\documentclass[twocolumn,10pt,letterpaper]{article}
\usepackage{usenix2019}
\usepackage{ifthen}
\usepackage{fancyhdr}
\usepackage{amsmath,amssymb}
\usepackage{xspace}
\usepackage{booktabs}
\usepackage{textcomp}
\usepackage{mathptmx}   % Times + Times-like math symbols
\usepackage{courier}
\usepackage{color}
\usepackage{graphicx}
\usepackage{multicol}
\usepackage{multirow}
\usepackage{adjustbox}
\usepackage{verbatim}
\usepackage{appendix}
\usepackage{titlesec}
\titlespacing*{\section}{0pt}{2.0ex}{1.0ex}
\titlespacing*{\subsection}{0pt}{1.5ex}{0.5ex}

\newcommand{\ANONAUTHORS}{Paper \#119}

\newcommand{\TITLE}{\system: Intra-Process Access Control for IoT Applications}

\newcommand{\PAGENUMBERS}{yes}

\newcommand{\ANONYMOUS}{no}
\newcommand{\SHOWTOAPPEAR}{no}
\newcommand{\COMMENTS}{yes}

\ifthenelse{\equal{\ANONYMOUS}{no}}{%
\newcommand{\AUTHORS}{\rm{\REALAUTHORS}}
\newcommand{\system}{Pyronia\xspace}
}{
\newcommand{\AUTHORS}{\ANONAUTHORS}
\newcommand{\system}{Pyronia\xspace}
}

\usepackage{mathptmx}

\usepackage[T1]{fontenc}
\usepackage[scaled=0.92]{helvet}
\DeclareMathAlphabet{\bm}{OT1}{ptm}{b}{it}

\newfont{\titlefont}{phvb8t at 17pt}
\newfont{\smalltitlefont}{phvb8t at 16pt}
\newfont{\authorfont}{phvr8t at 11pt}
\newfont{\affilfont}{phvr8t at 10pt}

\usepackage[absolute]{textpos}
\newcommand{\ToAppear}{%
\begin{textblock*}{\textwidth}(0.95in,0.4in)
\begin{flushright}
    \noindent{\fbox{\textsf{Draft version---please do not redistribute.}}}
\end{flushright}
\end{textblock*}
}

\ifthenelse{\equal{\PAGENUMBERS}{yes}}{%
\usepackage[nohead,centering,
            left=1in,right=1in,top=1in,
            footskip=0.5in,bottom=1in,
            columnsep=0.25in
            ]{geometry}
}{%
\usepackage[noheadfoot,centering,
            left=1in,right=1in,top=1in,
            footskip=0.5in,bottom=1in,
            columnsep=0.25in
            ]{geometry}
}

\usepackage{caption}
\usepackage{algorithm}
\usepackage[noend]{algpseudocode}
\newcommand{\Continue}{\State \textbf{continue} }

\usepackage{listings}
\newcommand{\URL}[1]{\url{#1}}

\newcommand{\eg}{e.g.,\xspace}
\newcommand{\ie}{i.e.,\xspace}

\newcommand{\Parabreak}{1.5ex}
\newcommand{\Paragraph}[1]{\vspace{\Parabreak}\noindent\textbf{#1}}

\newcommand{\ignore}[1]{}

\ifthenelse{\equal{\COMMENTS}{yes}}{%
\newcommand{\msm}[1]{\textbf{MSM: #1}}
\newcommand{\mjf}[1]{\textbf{MIKE: #1}}
} {
\newcommand{\msm}[1]{}
\newcommand{\mjf}[1]{}
}

\usepackage{enumitem}
\setlist[itemize]{noitemsep,nolistsep}
\setlist[enumerate]{noitemsep,nolistsep}
\usepackage{colortbl}
\definecolor{Gray}{gray}{0.9}
\definecolor{LightGreen}{rgb}{0.2,1,0.4}

\usepackage{longtable}

\usepackage{fancyhdr} 

\ifthenelse{\equal{\PAGENUMBERS}{yes}}{%
  \pagestyle{plain}
}{%
  \pagestyle{empty}
}

\date{}
\title{\TITLE}
\author{{\AUTHORS}%
\ifthenelse{\equal{\ANONYMOUS}{no}}{%
  \\*[0.1in] {Princeton University}}{}
}

\usepackage{microtype}  % SPACE

\begin{document}
\maketitle

%% Use the following at camera-ready time to suppress page numbers.
%% Comment it out when you first submit the paper for review.
\ifthenelse{\equal{\PAGENUMBERS}{yes}}{%
}{\thispagestyle{empty}}

\ifthenelse{\equal{\SHOWTOAPPEAR}{yes}}{\ToAppear}{}

\begin{abstract}
Third-party code plays a critical role in IoT applications,
which generate and analyze highly privacy-sensitive data. Unlike
traditional desktop and server settings, IoT devices mostly
run a dedicated, single application. 
As a result, vulnerabilities in third-party libraries within a process
pose a much bigger threat than on traditional platforms.
%Yet the only practical data protection tools available today 
%for IoT developers are not designed to prevent data leaks 
%from malicious or vulnerable third-party code \emph{imported} into an application.

We present \system, a fine-grained access control system for IoT
applications written in high-level languages. \system exploits developers'
\emph{coarse-grained} expectations about how imported third-party code operates
to
restrict access to files, devices, and specific network destinations,
at the granularity of \emph{individual functions}.
To efficiently protect such sensitive OS resources,
\system combines three techniques: system call interposition, 
stack inspection, and memory domains.
This design avoids the need for application refactoring,
or unintuitive data flow analysis, 
while enforcing the developer's access policy at run time.
Our \system prototype for Python runs on a custom Linux kernel,
and incurs moderate performance overhead on unmodified Python applications.

\end{abstract}

\section{Introduction}
\label{sec:intro}

Concerns about data security in the Internet of Things (IoT) have been
mounting in the wake of private data leaks in safety-critical technologies
(\eg \cite{ars-baby-monitor, jeep-hack, hospital-hack}) and large-scale
malware attacks exploiting vulnerable devices~\cite{symantec-iot-ddos,
dyn-hack, fridge-spam}. These concerns have driven application developers to
deploy measures that secure data in-transit to cloud
platforms~\cite{eclipse-iot-survey,aws-iot, nest-privacy}, or that detect
unauthorized code on devices~\cite{artik-protection, intel-iot}. %Yet, these
%ad-hoc safeguards do not prevent vulnerable or malicious third-party code
%\emph{within} an application from leaking sensitive information.

However, these safeguards cannot prevent vulnerable or malicious third-party
libraries \emph{within} IoT applications from leaking sensitive information.
Once a developer imports a vulnerable library, it runs with the
application's privileges and has full access to the application's resources
(e.g, files, devices and network interfaces). 
For example, a facial recognition library 
with a vulnerability in the \textit{recognize\_face()} function
could allow an attacker to steal the application's 
private authentication token by upload this file to an
unauthorized remote server. Meanwhile, the application developer
only expected \textit{recognize\_face()} to access the image 
file \texttt{face.jpg}.

This is an especially serious problem for IoT because most
devices run a single, dedicated application 
that has access to all sensors on the device. 
In other words, third-party code does not run with 
\emph{least privilege}~\cite{least-priv}.

Now, these threats are not IoT-specific. A large body of
prior research has sought to restrict untrusted 
third-party code in desktop, mobile, and cloud applications
(\eg~\cite{wedge,passe,breakapp,adsplit,flowfence,asbestos, 
histar, flume, tightlip, resin, hails,
cowl, aeolus, jif}).
However, these traditional compute settings face more complex 
security challenges.
Mobile devices, desktops, and even IoT hubs,
run multiple mutually distrusting applications; further,
desktops and cloud servers run in multi-user/multi-tenant settings.
All of these settings require isolation between the different 
applications and principals.

In this paper, we examine the following question (\S\ref{sec:related}): Are
traditional approaches suitable for protecting IoT device
applications against untrusted third-party code? 
Given the rapid proliferation of IoT devices and the high sensitivity
of the data they handle, it is crucial to gain an understanding of
the IoT-specific security challenges that developers face,
in order to guide the design of systems that effectively
enforce least privilege in this setting.

We conduct, to the best of our knowledge, the first in-depth
analysis of third-party code usage in IoT device applications.
Specifically, we characterize the IoT library landscape,
and identify key risks to IoT applications 
(\S\ref{sec:iot-app-analysis}). Informed by our findings,
we propose \system,\footnote{\system being a
gatekeeper for third-party libraries, is named after the genus of butterflies
known as gatekeeper butterflies.} an access control system
for untrusted third-party code in IoT applications written in high-level
languages.

In \system, we retain the goal of controlling when an application may obtain
data from files and devices, and to which remote network destinations an
application may export data. \system enforces a \emph{central} policy that
specifies rules for \emph{directly imported} library
functions and the specific OS resources they may access. For example, to
ensure that a sensitive image \texttt{face.jpg} is only accessible by the
\textit{recognize\_face()} function, the developer specifies a read rule for the 
\texttt{face.jpg} in her policy. \system then blocks all attempts by 
any other library functions to access the image file, as well as
attempts by this function to access other sensitive resources. 
Thus, developers need not reason about third-party
dependencies that may be unfamilar to them, and application and
library source code can remain unmodified.

To enforce such \emph{function-granular} access control,
\system leverages the following three techniques.

\begin{enumerate}[label=\textbf{\arabic*})]

\item \textbf{System call interposition} (\S\ref{secsec:stack-inspection})
guarantees that access to OS resources by all application components can be
controlled, even for native libraries integrated via a high-level native
interface such as the Java JNI or the Python C API. However, system call
interposition has traditionally only been realized as a process-level
technique, and thus cannot handle intra-process access checks.

\item \textbf{Stack inspection} (\S\ref{secsec:stack-inspection}) allows
\system to identify all library functions involved in the call chain that led
to a given system call. 
Thus, \system leverages the language runtime call stack to determine whether to
grant access to a requested resource based on the full provenance of the
intercepted system call.

\item \textbf{Memory domains} (\S\ref{secsec:memory-isolation})
are isolated compartments within a process address space,
each with its own access policy.
\system enforces boundaries between compartments via
replicated page tables, protecting the 
language runtime call stack against 
tampering by native code in the same address space.
\end{enumerate}

\system targets popular high-level IoT programming
languages like Java or Python~\cite{eclipse-iot-survey}
precisely for their ability to dynamically provide fine-grained
execution context information. We implement \system on Linux for Python 
(\S\ref{sec:implementation}), although we believe our approach can be applied to
other high-level languages.
Our prototype acts as a drop-in replacement for the CPython
runtime, and includes a custom Linux kernel with support for memory domains. 
Our function-granular MAC component is
a kernel module built on top of AppArmor~\cite{apparmor}. 
%To
%facilitate integration into a language runtime, we have implemented a \system
%userspace library, which provides APIs for loading a
%developer-supplied resource access policy, and for memory domain
%management.

We evaluated \system's security and performance with three open-source IoT
applications. Our security
evaluation (\S\ref{sec:use-sec-analysis})shows that \system mitigates reported and hypothetical OS
resource-based vulnerabilities. We also find that \system
incurs moderate performance slowdowns,
with a maximum operating overhead of 3x, and modest memory
overheads with an average of 38.6\% additional
memory usage for the entire application.
\section{Prior Work}
\label{sec:related}

Prior research on restricting untrusted third-party code and
enforcing least privilege in traditional compute settings
has used two main approaches. 

\Paragraph{Process isolation} partitions a
monolithic application into multiple processes and
controls their permissions individually (\eg~\cite{wedge,passe,breakapp,
adsplit,flowfence,codejail,adsplit}). 
However, process isolation imposes significant development and 
run-time overheads. Developers may have difficulty cleanly separating 
components into processes.
In addition, inter-process communication is much more expensive 
than calling library functions within the same address space.

These issues indicate that the process abstraction is too
coarse-grained for the efficient isolation of intra-process components
in IoT applications.
More recent work in this area
introduces OS-level abstractions to create private memory compartments within
a single process address space~\cite{lwc, smv, shreds, arbiter}.
However, these proposals lack built-in access control for OS resources,
and still require major developer effort for deployment in high-level applications.

Thus, these approaches are insufficient to limit third-party libraries'
access to unauthorized IoT device resources. Additionally, given the rapid deployment 
cycles of IoT applications, developers are unlikely to prioritize 
security~\cite{balebako,acar1,acar2}, and spend the time and effort necessary
to refactor their applications.
Thus, one key design goal of \system is to support unmodified applications
while enforcing least privilege within a single process.

\Paragraph{Information Flow Control}
attaches policy 
labels to individual sensitive data objects and 
track how they flow through the system. IFC systems can be divided
into two broad categories. OS-level IFC 
(\eg~\cite{asbestos, histar, flume, tightlip}) tracks
primitives like files or network sockets, while 
language-level IFC systems (\eg~\cite{resin, hails, cowl, aeolus, jif})
are capable of controlling access to sensitive data at
granularities as fine as individual bytes.

Yet, much like process isolation, IFC
systems introduce a considerable amount of cost and complexity: Developers
must manually refactor their source code to specify policy labels around
specific sensitive data sources and sinks. Additionally, propagating new
labels at run time incurs significant memory and performance overheads.

Since third-party code may
contain unexpected vulnerabilities, or a long list of dependencies,
we cannot expect IoT developers to be able to perform extensive
data flow analysis a priori to declare a data access policy that
minimizes data leaks. Thus, access control for IoT should
not require unintuitive policy specification.

%since we do not ex
%by transparently checking provenance information for syscalls, \system avoids
%the need for application developers to perform any unintuitive prior data flow
%analysis, or manual code annotations.

%Decentralized information flow control (IFC) systems explicitly label 
%data and track how they flow through the system, and can be divided
%into two broad categories. OS-level IFC systems 
%(\eg~\cite{asbestos, histar, flume, tightlip}) track
%primitives like files or network sockets, while 
%language-level IFC (\eg~\cite{resin, hails, cowl, aeolus, jif})
%is capable of controlling how entire variables to individual bytes
%flow through an application.

%Restrictions in IFC systems are usually 
%applied at disclosure time (e.g., filesystem write or network transmission),
%while no restrictions are imposed at read time. 
%For example, Asbestos~\cite{asbestos}
%allows any process to create new labels and raise the security clearance of
%other processes to facilitate reading of secret information. 
%\system does not track sensitive OS resources as they propagate 
%through the entire application, losing some protection granularity. However, 

%rather than only restricting where data may flow after writes.

\Paragraph{IoT-specific Access Control.}
A number of recent works in the IoT 
space~\cite{contexiot, eso, bochum-ac, tyche, smartauth} 
propose access control systems to
%to protect sensitive user data
%against potentially vulnerable or malicious IoT applications. 
%However, the majority of these systems focuses on
enable developers and end-users to define and enforce more suitable
data access policies based on 
external factors such as usage context or risk.
\system, in contrast, focuses on allowing developers to 
restrict third-party code decoupling end-user application usage policy
enforcement from a developer's implementation policy.

The FACT system~\cite{fact} 
%controls access to different
%IoT device functionalities by executing
%different IoT device functions in separate Linux containers.
%In contrast to \system, FACT 
aims to prevent overprivileged
\emph{applications} from accessing sensitive device functionalities and resources
by executing different IoT device functions in separate Linux containers.
However, FACT does not protect these resources against untrusted third-party code 
running as part of the isolated device functionalities, as \system would.
FlowFence~\cite{flowfence} shares the same goals as \system, 
but still relies on process isolation for in-application privilege separation,
and does not support unmodified applications.

\section{IoT Application Development Today}
\label{sec:iot-app-analysis}
%Before designing \system, we sought to understand the landscape of
%IoT application development.  
To design \system, we conducted an in-depth study of 85
open-source IoT applications written in Python and their libraries, as well
as a brief analysis of reported vulnerable Python libraries.
Our analyses focus on Python as it is a popular IoT
development language~\cite{eclipse-iot-survey}.

%\footnote{The IoT Developer Survey~\cite{eclipse-iot-survey} finds that Java
%is a more widely used language in the IoT, but third-party library usage for
%Java applications has been more extensively studied across platforms than for
%Python-based applications.}

\subsection{Application Analysis}
\label{sec:app-analysis}
To better understand how third-party libraries influence
IoT application design today, 
we analyzed 85 open-source IoT applications written in Python.
We obtained these applications primarily from popular DIY
web platforms such as \texttt{instructables.com} 
and \texttt{hackster.io}; our search focused on three broad categories of
applications---visual, audio, and environment sensing---which we
believe are representative of today's most privacy-sensitive IoT
use cases.
The four key takeaways of our analysis are:

\begin{table}[t]
\centering
\caption[Open-source IoT app dependency analysis.]{\label{tab:app-analysis} Analysis of direct imports and number of dependency levels in a set of 85 Python IoT applications and in the
top 50 third-party libraries imported by these applications. Unless noted otherwise, results of per-application analyses are shown.}
\begin{tabular}{ccccc}
\toprule
 & \textbf{min} & \textbf{median} & \textbf{mean} & \textbf{max} \\
\midrule
\# direct imports & 1 & 8 & 13 & 253 \\
\# direct 3p imports & 0 & 3 & 6 & 186 \\
\# dependency levels & 1 & 25 & 22 & 37 \\
\# lib dep levels & 0 & 27 & 24 & 34 \\
\bottomrule
\end{tabular}
\end{table}

\Paragraph{1. The vast majority of IoT applications import third-party libraries.}
%Most IoT products are single-purpose devices that run a single, dedicated
%application. This differs from traditional desktop or server settings where
%multiple application share the hardware of the same host. Although IoT devices
%are often single-purpose, 
All but one of our sampled applications (98.8\%)
import at least one third-party library, with a mean of about 6
\emph{direct} third-party imports per application. The maximum number 
of direct imports we found in a single application
was 186 (see Table~\ref{tab:app-analysis}).
This demonstrates that attacks from imported
libraries within an application's process are much more realistic 
due to the single-purpose nature of most IoT devices,
as opposed to threats across process boundaries as seen in traditional desktop and server settings.
%\footnote{It is worth noting that 
%imported libraries are typically not inspected by package managers (e.g., PyPI) for
%security.}

\Paragraph{2. The third-party library landscape is very diverse.}
Overall we found 331 distinct third-party Python libraries among
the 418 total imports in our sampled set of applications, despite
heavily sampling applications targeting 
the Raspberry Pi single-chip computer (a very
popular development platform for IoT).
Thus, providing intra-process access control
for IoT device applications requires an application- and
library-agnostic approach.
%, our sampled
%applications heavily target this platform.
%For instance, two of the top 3 most imported third-party libraries
%(\texttt{RPi.GPIO} and \texttt{picamera}) are Raspberry Pi-specific 
%Python libraries.% (see Table~\ref{tab:top5-libs}).
%Nonetheless, 

%\begin{table}
%\centering
%\caption[Top 5 Python IoT libraries.]{\label{tab:top5-libs} Five most popular Python IoT libraries, and
%the percentage of the 85 surveyed apps which included them.
%}
%\begin{tabular}{cc}
%\toprule
%\textbf{Library} & \textbf{Frequency} \\
%\midrule
%RPi.GPIO & 47.1\% \\
%requests & 23.5\% \\
%picamera & 21.2\% \\
%serial & 12.9\% \\
%alsaaudio & 11.8\% \\
%\bottomrule
%\end{tabular}
%\end{table}

\begin{table}[t]
\centering
\caption[Top 50 Python IoT library characteristics.]{\label{tab:top50-analysis} Characteristics of the top 50 IoT Python
  libraries, including libs exhibiting multiple characteristics.  }
\begin{tabular}{cc}
\toprule
\textbf{Library feature} & \textbf{\% of top 50 libs} \\
\midrule
Written in Python & 12.0\% \\
Have Native Deps & 82.0\% \\
Run external binaries & 40.0\% \\
Use ctypes & 40.0\% \\
\bottomrule
\end{tabular}
\end{table}

\Paragraph{3. Libraries rely predominantly on native code.}
%All of the third-party libraries in our study are Python libraries,
%\ie they provide a Python API.
We find that 82\% of the top 50 libraries 
in our analysis include a component written in C/C++, 
among which we identified 68 distinct native dependencies.
Additionally, a large portion of libraries load a native shared library via the \texttt{ctypes}
Python foreign function interface, or
execute external native binaries (including a Python subprocess).
These results reveal how exposed applications and the Python runtime are to threats from
vulnerable or malicious native code,
and underscore the importance of intra-process memory isolation to protect security-critical
data against tampering or leaks native code.

\Paragraph{4. Dependencies are nested dozens of levels in IoT applications.}
Finally, we analyzed the longest chain of nested
libraries for each application and top-50 library in our sample, 
and find that across the 85 sampled applications, the median number of nesting levels
is 25, while the median library has 27 levels of nested dependencies (see Table~\ref{tab:app-analysis}).
These numbers indicate the complexity of single applications and libraries,
and highlight why developers need more intuitive fine-grained access control that does not
require separating each library into its own process, or 
identifying the sensitive data flows within a single application.

\begin{table*}[t]
\begin{center}
\caption[Reported Python library vulnerabilities, 2012-2019.]{\label{tab:cve} Reported Python library vulnerabilities and number
of unique libraries by attack class, from 2012-2019.
}
\begin{tabular}{cccc}
\toprule
\textbf{Attack class} & \textbf{\# Reports} & \textbf{\# Libs} & \textbf{Lib/framework (\# reports)} \\
\midrule
Arbitrary code execution & 28 & 24 & python-gnupg (4) \\
MITM & 19 & 14 & urllib* (3) \\
Web attack & 18 & 12 & urllib* (4)\\
Denial of service & 17 & 12 & Django (3) \\
Direct data leak & 12 & 10 & requests (3) \\
Weak crypto & 11 & 10 & PyCrypto (2) \\
Authentication bypass & 9 & 6 & python-keystoneclient (3) \\
Symlink attack & 6 & 4 & Pillow (2) \\
Replay/data spoofing & 3 & 3 & python-oauth2 \\
\bottomrule
\end{tabular}
\end{center}
\end{table*}

\subsection{Library Vulnerabilities}
\label{secsec:lib-vulns}
Our survey of reported Python library vulnerabilities covers 123 reports
created between January 2012 
and March 2019 in the Common Vulnerabilities and Exposures (CVE) List~\cite{cve-python}.
We identify 78 distinct vulnerable Python libraries, and 9 main
attack classes (see Table~\ref{tab:cve})\footnote{For a full list of the CVE reports included in our analysis, see Appendix~\ref{sec:cve-list}.}.
%Our analysis does not include CVE reports for vulnerabilities in
%the cPython interpreter itself, as well as in Python-based applications.

We include shell injections under arbitrary code execution, and the 
majority of web attacks comprise cross-site scripting and CR/LF injection
attacks.
We classify vulnerabilities arising from accidental 
data exposure as direct data leaks.
Authentication bypass vulnerabilities arise
from system misconfiguration or credential verification bugs.
A symlink attack allows an adversary to
gain access to a resource via a specially crafted symbolic link.

While direct data leaks account for only about 10\% of the reported vulnerabilities,
we emphasize that 
%the actual number of reported data leak vulnerabilities is 
%significantly higher since 
most other attack classes, most notably
arbitrary code execution, man-in-the-middle (MITM), 
and authentication bypass, lead to information leakage as well.
Our analysis also demonstrates the diversity of vulnerable libraries, 
with a small number of libraries having a handful of reports in each attack class.
The two Python packages with the most overall CVE reports %in our survey
are the widely used Django web framework and urllib* HTTP library family, 
each with eight reports.
These findings underscore the degree to which IoT 
application developers are exposed to potential %security and privacy
threats by importing third-party code.

In addition, we find that exploiting dynamic language features is fairly straightforward.
In our lab setting, we built a Python module and a native library
that use %exploit dynamic language features of Python, such as 
introspection or monkey patching~\cite{monkey-patch}
to replace function pointers at run time with malicious functions, to leak
data at import time, and to perform various modifications to the contents of Python 
runtime's stack frames. 
%Importantly, these attacks are possible because Python exposes
%the exploited introspection APIs and features to provide programmers with 
%more flexibility, visibility, and control over a module's functionality.

While we have not identified such attacks in the wild, our experiments 
demonstrate that these 
dynamic features and open APIs place too much trust in third-party library developers,
and can be misused for nefarious purposes.
Thus, both dynamic language features and the capabilities of native libraries pose threats to the
integrity of the application itself and the privacy of user data.

%\Paragraph{Summary.} Third-party libraries pose a serious threat
%to sensitive IoT data.
%The single-purpose nature of most IoT devices means that attacks from imported
%libraries within an application's process are more realistic than those
%across process boundaries, as seen in traditional desktop and server settings.
%A system is needed to provide least privilege at the intra-process
%level to mitigate these threats. Crucially, this system must include defenses
%against native code to provide complete protection.

\section{Threat Model and Security Goals}
\label{sec:overview}

We aim to provide intra-process access control, 
which allows developers to prevent 
third-party code included in their IoT applications
from leaking data.
In particular, \system %aims to provide the comprehensive yet efficient
protects %protection of %two categories of 
sensitive OS resources, %and in-memory data objects
and restricts access to remote network destinations.

\subsection{Threat Model}
We assume that IoT device vendor, who usually also develop the
device software, 
are trusted. As such, we also trust the underlying device hardware, 
the operating system, and the language runtime executing the IoT application. 
Yet, imported third-party code poses a risk to developers:
library code is rarely inspected or readily accessible, 
so bugs or dynamic language features that 
leak sensitive data may go unnoticed.
%dynamic features of the Python language.
We do, however, assume that application developers do not \emph{intentionally}
include such vulnerable or malicious third-party code.%in their products. 
%However, third-party
%libraries that are imported by IoT developers are assumed to be malicious.
% \system tackles a strong adversary model:  
% MSM: this is more a general assumption and not part of the threat model.
% Fits better in design.
%We do assume that application developers understand
%the purpose of imported libraries and can provide high-level descriptions of
%their expected resource access behaviors.

%We emphasize that \system's threat model applies most directly to IoT
%despite the fact that these threats are not IoT-specific. 
%The majority of IoT devices are single-purpose running a
%dedicated application, with the primary threat stemming from third-party
%libraries. Traditional compute settings (\ie mobile, desktop, cloud),
%on the other hand, face more complex security challenges. 

While data leak vulnerabilities take many forms, \system targets
third-party code that aims to access arbitrary sensitive
files or devices, or exfiltrate sensitive data to arbitrary remote
network destinations. 
\system does not seek to prevent any control flow (\eg ROP~\cite{rop}) or 
side channel attacks (\eg Spectre/Meltdown~\cite{spectre,meltdown}, or physical vectors~\cite{ecdsa-phys-sidechan}). 
%Control flow attacks such as ROP attacks pose a challenge to \system's
%memory isolation. 
%Like previous proposals,
ROP defenses (\eg~\cite{ropdefender,binary-stirring,uct,ropecker}) may be used in a complementary fashion. 
\system also does not prevent network-based attacks such as man-in-the-middle
or denial-of-service attacks.

\subsection{Security Properties}
%To maintain strong data protections in the face 
%of buggy or vulnerable third-party code,
\system provides three main security properties.

\Paragraph{P1: Least privilege.} A third-party library function may 
only access those OS resources (\ie files, devices, network)
that are necessary to provide the \emph{expected} functionality.
Attempts by a third-party function to access resources that are not
relevant to its functionality must be blocked.
\system conservatively enforces a default-deny 
policy, 
requiring developers to explicitly grant specific library functions access to OS resources. 

%\Paragraph{P2: Data isolation.} Sensitive in-memory data objects
%may only be accessed within the scope of those third-party library functions
%which are \emph{expected} to operate on them.
%That is, a third-party function cannot read or modify
%any protected data objects unless it is called by an authorized library
%function. \system enforces this property by placing sensitive in-memory
%objects (including the language runtime call stack), into strongly isolated
%memory domains. Access to these memory domains is only granted by the language
%runtime upon entry to an authorized third-party function, and is revoked after
%the function returns.

\Paragraph{P2: No confused deputies.} All access control decisions
are made based on the \emph{full provenance} of the access request.
This prevents confused deputy attacks~\cite{confused-deputy} in which an
underprivileged library function attempts to gain access to a
protected resource via another function that does have sufficient privileges. 
To detect such attempts to bypass access control checks, \system 
%leverages the language runtime call stack to 
checks all functions involved
in a request for an OS resource.

%\Paragraph{P2: Data isolation.} Sensitive in-memory data, such as 
%security-critical language runtime data, may only be accessed
%by authorized in-application functions.
%%That is, a third-party function can  directly modify
%%any sensitive security-critical metadata. 
%\system places 
%sensitive data in strongly isolated memory compartments within the application's
%address space. 
%%Access to these memory domains
%%is only granted upon the creation of a new stack frame
%%or local variable modifications, and is revoked after these functions return.

\Paragraph{P3: Verified network destinations.} A third-party library function
may only transmit data to remote network destinations
(\eg cloud servers or other IoT devices) whitelisted by the developer.
Thus, a third-party library cannot leak legitimately collected data to
an untrusted remote server or device.
\system prevents such data exfiltration by intercepting all outgoing network
connections.
%and  of the network connection request. 
%Then, \system verifies that all 
%involved functions have sufficient privileges to transmit data to the requested
%network destination.

\Paragraph{Non-goals.} While \system automates
access control at the level of in-application components,
our design does not seek to provide automated execution isolation 
of these components (\eg~\cite{breakapp, passe, wedge}).
We also do not guarantee the \emph{correctness} of the sensitive data they
output. Automated code compartmentalization is complementary to our approach, and
could be added to \system to allow developers to prevent certain cross-library
function calls. Ensuring the correctness of sensitive outputs, on the other
hand, could provide additional data leak protection.
% for cases in which a 
%vulnerable or malicious library designed to process sensitive data simply returns the
%unmodified input, or embeds other sensitive information as part of its output.
However, formally verifying the functionality of untrusted code is beyond the scope of
\system, and could be performed separately prior to application
deployment.

%\Paragraph{Deployability Goals}
%\begin{itemize}[label={}, leftmargin=0pt]
%\item \textbf{G4: Platform-independence.} Because of the heterogeneity of IoT software and hardware
%platforms, \system aims to be compatible with multiple
%popular IoT implementation language runtimes as well as any Linux-based platforms. 
%\item \textbf{G5: Unmodified libraries and applications.} To allow for a general approach and 
%to reduce the burden on developers, \system does not require modifications to 
%third-party library or application source code.
%\item \textbf{G6: Efficiency.} The computational and memory overhead
%of using \system should be minimal, 
%so that it is feasible for \system to 
%protect IoT applications running on a broad range of devices
%and platforms without substantially affecting their performance.
%\end{itemize}

\section{System Design}
\label{sec:design}

\system enforces %\emph{end-to-end} 
intra-process least privilege
without partitioning an application into multiple processes,
or propagating data flow labels.

Developers understand the purpose of imported libraries and 
can provide high-level descriptions of their expected data access behaviors.
\system thus relies on developers to specify all access rules in a single, central policy
file. 
At run time, \system loads this file into an application-specific 
access control list (ACL) that contains an entry for each developer-specified
library function and its associated data access privileges.
\system imposes \emph{default-deny} access control semantics,
meaning that a third-party library function may only access
those files, devices, % in-memory data objects, 
and whitelisted remote network destinations.

\begin{figure}[t]
  \centering
    \includegraphics[width=0.4\textwidth]{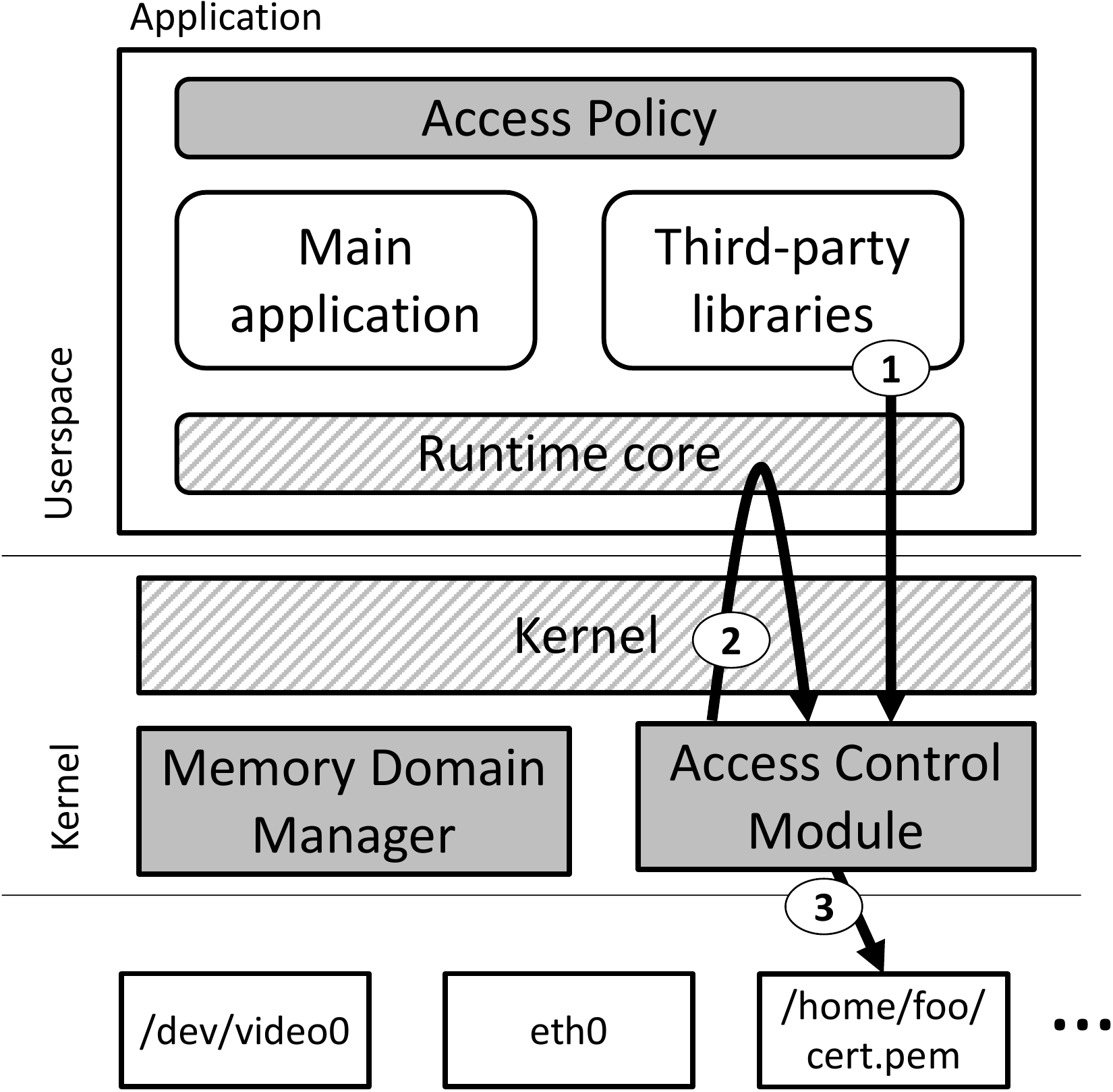}
    \caption[\system system architecture.]{\label{fig:sysarch} Overview of \system, which enforces function-granular resource access policies via runtime
      and kernel modifications (striped boxes). New features 
      are represented by gray boxes. The arrows show the components involved
      in an access request to a file such as a certificate.}
\end{figure}

To enforce this policy securely, \system requires support both 
in the language runtime and from the OS.
Figure~\ref{fig:sysarch} provides an overview of the \system system
architecture, and shows the main steps involved in a resource access request
(see~\S\ref{secsec:stack-inspection}).

\subsection{Function-granular MAC}
\label{secsec:stack-inspection}

At first glance, performing access control at library function granularity
in the language runtime may seem sufficient. The
runtime can directly inspect its function call stack when a specific 
third-party library function
uses the language's high-level interface to access a file or device. If the
a function with insufficient permissions attempts to access a resource, 
the runtime can block
the request and throw an error to notify the application.

However, language runtimes also provide an interface to native code, such
as Java's JNI or Python's C API;  indeed, our analysis in~\S\ref{sec:app-analysis}
shows that use of this interface in Python is very
common-place. This ability to include native code in otherwise memory-safe
languages exposes applications to vulnerabilities in native code:
as this code runs outside the purview of the language runtime, this code could
bypass any runtime-level access control via direct system calls.

\system addresses these issues via system call interposition (\eg~\cite{janus,systrace,interposition-agents})
enhanced with function call provenance. Many mandatory access control (MAC)
systems in deployment use system call interposition (\eg SELinux~\cite{selinux},
AppArmor~\cite{apparmor}, Windows MIC~\cite{windows-mic}). Their limitation,
however, is that the security policy is only enforced at the process level.
\system achieves intra-process policy enforcement by incorporating
runtime call stack inspection
(\eg~\cite{flexdroid, compac, quire, java-stack-inspection})
to obtain the full provenance of the system call.
%the most effective technique known today for mediating access to OS resources.
%because even native code within a language runtime cannot bypass system calls.

With these two techniques, \system provides
\emph{function-granular} MAC to enforce least privilege (\textbf{P1}).
As we show in Fig.~\ref{fig:sysarch}, when the application attempts to 
access a sensitive OS resource (\eg an SSL certificate), the \system Access Control module in the 
kernel intercepts the associated system call (step 1).
This kernel module then sends a request to the language runtime via a 
trusted \emph{stack inspector} thread, which pauses 
the runtime. After collecting the interpreter's function call stack 
in its current state, the stack inspector returns the stack to the kernel (step 2).

\begin{algorithm}[t]
\caption{Call stack inspection}\label{alg:si}
\begin{algorithmic}[1]
\Procedure{InspectStack}{$callStack$, $reqAccess$}
\State $rule \gets \Call{GetAclRule}{callStack.funcName}$
\If{$rule ==$ nil}
	\Return false
\EndIf
\State $grantAccess \gets \Call{HasPrivs}{rule, reqAccess}$
\While{$grantAccess ==$ true}
	\State $callStack \gets callStack.Next$
	\State $rule \gets \Call{GetAclRule}{callStack.funcName}$
	\If{$rule ==$ nil}
		\Continue
	\EndIf
	\State $grantAccess \gets \Call{HasPrivs}{rule, reqAccess}$
\EndWhile
\Return $grantAccess$
\EndProcedure
\end{algorithmic}
\end{algorithm}

The Access Control module maintains the ACL for
the developer-supplied policy.
To determine whether to grant access to the requested OS resource,
the module inspects the call stack to verify the provenance
of the system call. Only if all developer-specified functions identified in the
call stack have sufficient privileges may the
application obtain data from the requested resource (step 3). 
That is, to determine the application's access permissions
to the requested resource, \system dynamically computes the intersection of the
privileges of each function in the call stack using algorithm~\ref{alg:si}.
This algorithm prevents confused deputies (\textbf{P2}), 
much as in~\cite{flexdroid,quire,ipc-inspect}.

\subsection{Runtime Call Stack Protection}
\label{secsec:memory-isolation}

Untrusted native libraries reside in the
runtime's address space, giving them unfettered access to the call stack's
memory region. A malicious native library may tamper with the runtime call stack
in an attempt to bypass \system's function-level MAC.

This challenge is not unique to \system; indeed, prior work in the mobile 
space~\cite{flexdroid, compac} recognized the need to protect the Dalvik call
stack against native third-party code in the trusted host app's address space.
To address this issue, these proposals either rely on special hardware 
support~\cite{flexdroid} to separate the runtime address space from the 
native library address space, or they forgo memory protection altogether~\cite{compac}.

\system, in contrast, aims to provide a more generally applicable solution
to this issue, since IoT software runs on a very diverse range of hardware platforms.
We overcome this challenge with page table replication, a technique that enables us to
create strongly isolated
memory regions, or \emph{memory domains}, within a process' address space.
Prior work in this space (\eg~\cite{arbiter, smv, lwc}) has introduced new
primitives for intra-process execution compartments. 
Our design for \system's memory domains, 
on the other hand, focuses on data isolation.

To protect the language runtime call stack for a broad range of
applications, 
memory domains in \system
meet two requirements: (1) the size of a domain must be flexible, and 
(2) the access privileges must be dynamic. The first requirement is important
for ensuring that \system can support applications that make an arbitrary number of nested function calls. 
The second requirement allows \system to restrict an application's access
to a memory domain at run time based on the currently executing code (\eg interpreter versus
third-party library), while still enabling data sharing between application
components. % that may need access to the protected data at different times.

The \system Memory Domain Manager in the kernel (see Fig.~\ref{fig:sysarch})
maintains a per-process table of domain-protected memory pages,
and replicates the corresponding page table entries (PTEs); each replicated entry 
is associated with a distinct domain access policy. 
Policies are enforced at the granularity of individual native threads, and specify
the access permissions for any thread launched under a given policy. 
Mapping thread contexts to a replicated PTE thus allows 
\system to transparently change policy contexts during a context switch.

Upon a memory access, the \system kernel 
performs all regular memory access checks. 
If the requested address
is domain-protected, the Domain Manager additionally
verifies that the loaded thread context has sufficient permissions to
access the requested memory domain based on the thread's policy. 
Any attempt by an application to access
unauthorized domain-protected memory results in a memory fault.

To ensure the integrity of the runtime call stack, the \system runtime allocates
all call stack data into a memory domain called the \emph{stack domain}.
\system currently defines two access policies to this domain:
The runtime's policy, which only allows access to the call stack
during stack frame creation and deletion, and the stack inspector's policy,
which allows access to the call stack while responding to an upcall from
the Access Control module.
\system loads the appropriate page table during context switches between
the main runtime thread and the stack inspector thread automatically. 

However, to enforce the runtime's policy, \system requires
elevated access privileges to the stack domain during
stack frame creation and deletion.
Thus, the runtime invokes the Domain Manager to temporarily adjust the main
runtime thread's policy and corresponding PTE access bits during these operations.
%Upon completion of these operations, the \system runtime disables write access to
%the interpreter domain to protect the call stack. 
Otherwise, the language runtime still provides 
\emph{read} access to the stack domain to allow code to
make use of shared functions, but prohibits \emph{write} access to protect
this security-critical metadata.

We note that while our memory domain mechanism
can support multi-threading beyond \system's threads,
\system currently only targets single-threaded applications.
As we describe in~\S\ref{sec:discussion}, we leave this enhancement
as future research due to the small fraction of IoT applications
(10\% per our analysis in~\S\ref{sec:app-analysis})
that spawn multiple threads.

\subsection{Network Destination Verification}
\label{secsec:network-dest-verif}
Since all IoT applications communicate with
remote services, \system must ensure that authorized third-party library
functions may transmit data only to \emph{whitelisted} destinations.
That is, when an application attempts to export data via
the network, \system's Access Control module intercepts
all outward-facing socket system calls 
(\eg \emph{bind()} and \emph{connect()}) for all socket types, \ie
TCP, UDP, and raw.
As with other OS resource accesses, \system then
requests and inspects the runtime call stack.

However, network access privileges alone do not immediately
allow a third-party function to transmit data. \system also
verifies the remote endpoint for the requested socket.
Thus, only if the address of the requested destination is whitelisted
for the given third-party function, does \system grant access to the
requested socket (\textbf{P3}).

\subsection{Child Process Protection}
\label{secsec:child-proc-continuity}
We find in ~\S\ref{sec:app-analysis} 
that a large fraction of applications (46\%) and libraries (40\%)
run external binaries (including new language runtime instances)
in subprocesses.
This application characteristic poses a challenge to \system's
function-granular MAC since child processes run in an
independent context, losing important provenance information
for system calls.
\system addresses this issue by ensuring the continued protection
of child processes. Thus,
upon a \texttt{fork()} system call, the \system kernel propagates
the parent's process- and function-level ACLs, as well as the
replicated page tables, to the child process.

However, the Access Control module does not request the current call
stack from the language runtime at a \texttt{fork()} for three reasons. First, the 
parent process may be a native binary that is \system-unaware. Second, since
application developers may be unaware of subprocesses spawned by third-party code, 
using the call stack to block external binaries may break the functionality
of the overall application. Third, using the parent
runtime's call stack for access control decisions in its children
may also unduly restrict the functionality of the whole application.
Thus, \system requires language runtime sub-instances to register themselves
with the \system kernel to transparently enforce the developer's
function-level access policy, and to continue managing the 
stack domain in the child's address space. In the case of a native child process,
\system does not provide intra-process access control, 
but still enforces the developer's policy
at process granularity.

\section{Implementation}
\label{sec:implementation}

To demonstrate how \system interacts with 
existing open-source IoT applications, we implemented the 
\system kernel based on Linux Kernel version 4.8.0+, and 
the \system runtime as a modified version of Python 2.7.14.
We have released all components of our prototype
on Github.\footnote{\label{foot:github}https://github.com/pyroniasys}

\subsection{Policy Specification}
\label{secsec:policy-spec}

\begin{table*}[t]
\centering
\caption[\system policy specification.]{\label{tab:rule-formats} \system data access rule specification formats. Supported access privileges are read-only $r$, and read-write $w$. }
\begin{tabular}{cc}
\toprule
 \textbf{Rule type} & \textbf{Format} \\
\midrule
FS resource & \small{\texttt{<module>.<function name> <path to resource> <access privs>}} \\
%Data object & \small{\texttt{<Python module>.<function name>:<object label> <access privileges>}} \\
Network destination & \small{\texttt{<module>.<function name> network <IP addr/prefix>}} \\
\bottomrule
\end{tabular}
\end{table*}

Application developers in \system specify all function-level data access rules
for file system resources, and network destinations in a single policy file.
Table~\ref{tab:rule-formats} details \system's policy rule specification
format.
% that is flexible capturing all three access rule types and different
%access privileges, while remaining concise to allow for single-line rules.
In the case of network access rules specifically, our \system 
prototype allows developers to specify
IP address prefixes, as the
specific address of a remote destination in a cloud service
may not be known a priori.

Nevertheless, a challenge that arises from having limited knowledge
about a library's implementation is that it may legitimately require
access to other resources that are unexpected. 
For example, a library function may need 
access to system fonts, or may write an intermediate result
to the file system.
Similarly, developers are unlikely to have a good sense 
of the system libraries or other file system locations a language 
runtime requires to operate properly.
Thus, \system's default-deny access control semantics alone
are too restrictive and may lead to a number of false negatives.

%One approach to address this challenge would be to provide a
%testing mode for \system, in which developers obtain a log of
%all requested resource accesses, akin to proposals like~\cite{systrace, passe}.
%Yet, we cannot expect developers to have sufficient understanding
%of the requested resources, nor the ability and patience to
%examine these logs to build a comprehensive policy.

To maintain the functionality
of the application and the language runtime, 
and reduce the number of false negatives,
our prototype supports a special 
\emph{default} access rule declaration, which grants application-wide
access to a specified resource. While default access rules bypass
\system's function-level access control, they still provide
a baseline level of security as \system also applies
default-deny access control semantics at the process-level.
%that preserve the functionality
%of the \system runtime and most applications, in the face of such 
%execution ``side-effects''. 
%In particular, based on three open-source IoT applications we evaluated (see~\S\ref{secsec:apps}), we identified 
%a set of default system files and libraries (\eg the \texttt{\textbackslash{}etc\textbackslash{}hosts} file, or \texttt{libdl.so}).
%These system files are required for the Python runtime itself,
%and hence may not be accessed as part of Python code execution.
%In other words, a runtime call stack may not exist at the time
%of access.

%To allow the interpreter to access these system files
%at any point, 

%Our prototype requires that all default access rules be
%individually specified in the developer's policy. However, 
%we facilitate policy specification by providing a
%policy generation tool that creates a template file that
%includes a number of pre-identified Python runtime default system files.
%A future iteration of \system could maintain a list of the most
%critical default rules within the kernel,
%allowing developers to remain agnostic to the runtime defaults.
%
\subsection{\system Kernel}
\label{secsec:kernel}

\begin{algorithm}[t]
\caption{\system in-kernel access control check}\label{alg:stack-log}
\begin{algorithmic}[1]
\Procedure{CheckAccess}{$acl$, $reqAccess$}
\State $grantAccess \gets$ true
\State $isDefault \gets \Call{IsDefaultAccess}{acl.resource}$
\If{$isDefault ==$ true}
	\Return $grantAccess$
\EndIf
\State $recvHash \gets \Call{RecvStackHash}{acl.resource}$
\State $hashLog \gets \Call{StackHashLog}{acl.resource}$
\If{$recvHash \neq$ null AND \\
	\hskip\algorithmicindent $hashLog.\Call{Contains}{recvHash} ==$ true}
	\State \Return $grantAccess$
\EndIf
\State $callStack \gets \Call{GetStackFromRuntime}{\null}$
\State \Return $\Call{InspectStack}{callStack, reqAccess}$
\EndProcedure
\end{algorithmic}
\end{algorithm}

\Paragraph{Access Control module.}
Our Access Control Module extends the AppArmor~\cite{apparmor} kernel module
version 2.11.0. 
AppArmor interposes on all system calls enforcing a process-level access policy.
Thus, as in vanilla AppArmor, \system denies access to a requested resource
if the \emph{process} does not have sufficient privileges. 
To add support for \system's stack inspection, we extend the process-level 
AppArmor policy data structure with a function-level ACL.
This ACL is populated at application initialization (see~\S\ref{secsec:python}),
and contains an entry for each developer-specified library function.
In addition, the Access Control module registers child processes
of the main \system process, propagating the application's function-level ACL
to all child processes.

In an early version of our prototype, \system would inspect the runtime call stack
on \emph{every} intercepted system call. However, given that IoT applications
are built to continuously gather and transmit data in an infinite loop, we 
improve the performance of \system's function-level MAC (by roughly 3x) by avoiding
expensive kernel-userspace context switches for already-verified call stacks.
Thus, we implement a \emph{stack logging} mechanism, which stores and verifies
the SHA256 hash for up to 16 functions authorized to access a given resource,
as part of the function-level access control checks, 
outlined in Alg.~\ref{alg:stack-log}.

The Access Control module first checks the 
defaults ACL for the application (line 3).
If the requested resource is covered by a default rule, \system
grants access without inspecting the call stack.
Otherwise, the Access Control module checks whether the 
the Python runtime has included a call stack hash
along with the system call (line 5, more details in~\S\ref{secsec:python}).
If kernel received a call stack hash, and 
the received hash matches any of the hashes in the call 
stack log for the requested resource (line 8), 
the module grants access. 

Otherwise, our prototype resorts to the full stack inspection mechanism
(lines 9 and 10, see Alg.~\ref{alg:si}).
Once the call stack has been inspected, and if the application has sufficient
privileges to access the requested resource,
the Access Control module logs the SHA256 hash of the 
callstack in the resource's ACL. 
To support this mechanism, we made minor, backwards-compatible
modifications to the
the \texttt{open()} and \texttt{connect()} system call code in order to
to parse the received call stack hash, if any, and store it 
for later verification during the MAC checks.

%If the
%requested resource has been given default access, 
%the Access Control module permits the system
%call to complete without requesting the runtime call stack.
%However, if the requested resource does not have default access,
%the module sends a call stack request to the stack inspector thread running in the
%language runtime. 
%After receiving the call stack, the Access Control module checks the call stack
%as detailed in~\S\ref{secsec:stack-inspection}. 

\Paragraph{Memory Domain manager.}
For \system's page table replication, we modify the SMV~\cite{smv}
kernel module, a memory isolation proposal
for enforcing per-page access control
policies via thread-local page tables. 
%While the main goal of SMV is to provide
%isolation for multi-threaded applications, this system
%still offers the appropriate abstractions to implement \system 
%memory domains. 
Our Memory Domain manager leverages the
SMV kernel API for maintaining a list of protected 4-kB domain
pages and their corresponding access policies for the
main runtime and stack inspector threads.
Upon a \texttt{fork()}, the kernel copies the replicated page tables
in the child process.
%(in \system, the main and stack
%inspector threads, each with their own domain access policy). 

\Paragraph{Netlink sockets.} To enable communication between the kernel
and the runtime in userspace, \system uses a generic Netlink socket
in the Domain manager, and one in the
Access Control module. Netlink sockets offer two
advantages: (1) they allow bi-directional communication between
kernelspace and userspace obviating the need to implement
additional ioctls() or system calls, and (2) userspace applications
can use the POSIX socket API for Netlink communication. 
%regular socket-based communication.

\subsection{\system Python Runtime}
\label{secsec:python}
To allow developers to run completely unmodified 
applications, the \system runtime acts as a drop-in replacement
for Python. We integrate our \system library, which provides an API for 
loading the developer's access policy, and for managing the stack memory domain. 

\Paragraph{Policy initialization.} The runtime core uses our policy parser API 
to read the developer's policy file during interpreter initialization. 
All parsed OS resource and network access rules are sent to
the Access Control module in the kernel. %initializing the function-level ACL for the application.
%The in-memory object access rules, as well as the object label-to-domain
%mappings, are stored as part of the runtime's \system security context.
Loading the policy before the runtime has loaded any third-party code
has the advantage of preventing an adversary from ``front-running'' the
interpreter by initializing the application's in-kernel \system ACL before
the legitimate developer-supplied policy can be loaded.
For this reason, the \system runtime also spawns the stack inspector thread
and registers it with the kernel during the initialization phase.

\Paragraph{Stack domain allocation.}
As the userspace \system memory domain management API
acts as a drop-in replacement for malloc, we instrumented the 
Python stack frame manager to allocate new runtime call stack
frames in the \emph{stack domain}. % to use \system memory domain allocation.
Because write access to the stack domain is disabled by default,
the runtime temporarily obtains write access to this domain during
frame creation and deletion operations.

\Paragraph{Child processes.}
Our \system runtime provides continuous protection for child 
processes spawned via standard Python APIs (\eg \emph{os.system()}).
%To provide this continuity, the \system initialization function keeps track of 
%the main application process' PID, in order to detect when the runtime
%forks a child process. 
As forking preserves the parent process' memory in the child,
\system subprocesses automatically inherit the parent's memory domain
layout as well as the \system interpreter metadata, including currently writable
memory domains.
Thus, \system initialization in child processes only requires spawning the child's
stack inspector thread, and resetting the application's access permissions
to the stack domain disabling write access to this domain. This reset is necessary
to ensure that the child cannot access any runtime stack frames in its own
address space that the parent process had marked as writable at the time
of forking.

%Importantly, initialization of \system occurs automatically in child processes
%forked using standard Python APIs such as \textit{os.system()}, allowing
%the continued protection of child processes without developer intervention.
%
\Paragraph{Stack logging.}
As described in~\S\ref{secsec:kernel}, we implement a stack logging
mechanism to reduce the overhead of kernel upcalls for runtime call stack requests.
When the language runtime requests
a resource for the first time, the \system performs the full stack inspection
mechanism. If the kernel grants access to the requested resource,
the \system runtime logs the resource as authorized. Then, in subsequent 
system calls, our prototype first checks this log; 
if the requested resource has been logged,  
the runtime preemptively collects its current call stack \emph{before} making the
system call, computes the SHA256 hash,
and embeds this hash into the input to the upcoming syscall.

To enable this optimization, we created system call-specific wrappers
(via LD\_PRELOAD) for 
various variants of \texttt{open()} and \texttt{connect()}.
These wrappers perform the preemptive call stack collection, 
hash serialization, and authorized OS resource logging. 
These wrapper functions are backwards-compatible, and do not
affect function-level policy specification.

\Paragraph{Garbage collection.}
Object reference counting used for Python's garbage collection
poses a challenge to protecting the stack domain without
breaking the functionality of the application.
Specifically, we found that Python increments or decrements 
several objects' reference counts, including
those of stack frames, for practically every Python instruction 
and internal operation.

To address this issue,
%without unduly weakening data isolation,
we temporarily elevate the interpreter's permissions to the stack domain %and object domain
around those blocks of
the Python runtime code that operate on domain-protected data.
Yet, simply granting write access to the entire stack domain
is inefficient, %especially for the stack domain, which 
since it may cover hundreds of pages, each of whose page table
entries would need to be modified (requiring TLB flushes at high rates).

%\system's memory domain implementation provides us
%with an opportunity to optimize these frequent domain
%access adjustments.
We optimize these frequent domain access adjustments by 
tracking the addresses of stack domain pages in the \system runtime,
%Because memory domains enforce per-page access control,
%we can greatly reduce the number of domain pages
%that need to be adjusted if we can determine which
%individual domain pages need to be accessed in a given interpreter
%code block.
%we can limit each adjustment to a single page
%given the address of the stack frame or other domain-protected
%metadata %in-memory object 
%that the interpreter needs to access.
%To this enable this optimization, the \system runtime 
%thus keeps a list of pages in the stack domain,
and only modifying the access privileges to a specific page when needed.
One exception to this is for new stack frame allocations: 
since the runtime is creating a new buffer with an undetermined
memory address, our prototype enables write access to all domain pages
with free memory chunks. %before these operations.
%and identifies which domain pages must be adjusted based
%on the address of the stack frame or in-memory object that the 
%interpreter is modifying.
\section{\system Protection in Real Applications}
\label{sec:use-sec-analysis}

To examine the effectiveness of \system's policies
and protections in real applications, we conduct three in-depth case studies
of Python applications that specifically capture a range of common IoT use cases,
and import a variety of Python libraries.
Our goal is to answer the following questions about the usability and security
of \system: \\
\textbf{\S\ref{secsec:policy-usability}} How difficult is it to write function-level access policies for \system? \\
\textbf{\S\ref{secsec:sec-analysis}} What reported Python library vulnerability classes can \system mitigate? \\
%\textbf{\S\ref{secsec:circumvention-analysis}} Is \system robust against attempts to circumvent function-level MAC? \\
\textbf{\S\ref{secsec:dyn-analysis}} What effect do dynamic language features of Python have on \system's protections?

\subsection{Case Studies} 
\label{secsec:apps}

We evaluate three open-source Python IoT applications that 
represent the main categories of applications we studied: visual, audio, 
and environmental sensing.
Each of these applications communicates with a cloud service for data processing or
storage, which required that we register an account to obtain authentication credentials.
The imported libraries in our study implement a broad range of 
common IoT functionalities (see Table~\ref{tab:app-policies}). Specifically, 
our goal is to study how \system operates for common IoT authentication 
mechanisms, data processing techniques, and communication protocols.

Through manual inspection of the source code of each case study, 
we found four distinct direct \emph{third-party} imports,
and three standard libraries.
While small in number, these direct imports are among %the top 50 overall or
top 50 third-party imports in our analysis in \S\ref{sec:app-analysis}.
%Table~\ref{tab:app-policies} gives an overview of our case studies.

%\begin{table*}[t]
%\caption{\label{tab:app-policies} Summary of case study applications.}
%\centering
%\begin{tabular}{ccp{2in}c}
%\toprule
% \textbf{App} &  \textbf{3p imports} & \textbf{Highlighted features} & \textbf{\# func-level rules}\\
%\hline
%\texttt{twitterPhoto} & tweepy & integrated OAuth, HTTP & 5  \\
%\midrule
%\texttt{alexa} & json & data marshalling & 0 \\
% & memcache & raw sockets & 1 \\
% & re & regex parsing & 0 \\
% & requests & widely used HTTP API & 10 \\
% \midrule
%\texttt{plant\_watering} & paho-mqtt & MQTT comm, binary exec & 11 \\
% & ssl & crypto, native dependencies & 0 \\
%\bottomrule
%\end{tabular}
%\end{table*}

\begin{table}[t]
\caption{\label{tab:app-policies} Summary of case study applications.}
\centering
\begin{tabular}{ccc}
\toprule
 \textbf{App} &  \textbf{imports} & \textbf{Highlights}\\
\hline
\texttt{twitterPhoto} & tweepy & integrated OAuth \\
\midrule
\texttt{alexa} & json & data marshalling \\
 & memcache & raw sockets \\
 & re & regex parsing \\
 & requests & HTTP API \\
 \midrule
\texttt{plant\_watering} & paho-mqtt & MQTT, binary exec \\
 & ssl & crypto, native deps \\
\bottomrule
\end{tabular}
\end{table}

\Paragraph{\texttt{twitterPhoto}} takes an image from
a connected camera every 15 minutes, and sends the picture
along with a short message to a specified Twitter account.
Before sending the tweet, the application authenticates itself
to Twitter via OAuth. The tweepy library is used both to
authenticate the app and upload the message. 
%We obtained this application from the set of
%85 applications described in~\S\ref{sec:app-analysis}.

\Paragraph{\texttt{alexa}} provides an open-source Python 
implementation of an Amazon Echo smart speaker. This application 
records audio via a microphone while a button is pressed, 
and sends the recorded audio (along with
authentication credentials) to the Alexa Voice Service 
(AVS)~\cite{alexa-voice} for processing.
The AVS sends an audio response, if the recorded
data is one of the commands recognized by the service, for
the \texttt{alexa} application to play back. Otherwise,
the AVS responds with an empty ACK message. 
%in which case the application does not play back
%any audio.
%
The python-memcache library is used to cache the app's AVS
access token, and the requests library for communicating with the
AVS via HTTP. This app also uses the json library to format all
messages exchanged with AVS, and the re library to 
parse out any audio file contained in the AVS response.
To facilitate our security and performance tests, 
we removed the button press
for audio recording, and instead open a pre-recorded audio file. 
%This application is also among the 85 applications we analyze in~\S\ref{sec:app-analysis}.

\Paragraph{\texttt{plant\_watering}} records moisture
sensor readings once a minute, and sends them to the Amazon AWS IoT~\cite{aws-iot-service}
service via MQTT~\cite{mqtt}, a widely used IoT communication protocol. 
MQTT also handles client authentication with the AWS IoT service via TLS.
We replaced sensor readings
with a randomly generated value to facilitate testing,
%We adapted this application from an AWS IoT SDK
%tutorial~\cite{plant-watering-tutorial}; 
and replaced the original MQTT library with one that supports single-threaded
network communications.
%, in place of the original application's use of
%the AWS IoT Python SDK~\cite{aws-iot-sdk}, which only supports a multi-threaded 
%MQTT client.

\subsection{Specifying Function-Level Policies}
\label{secsec:policy-usability}

To understand whether \system's policy specification
places an undue burden on developers, we analyzed the policy specification
process for our three case studies.

As we describe in~\S\ref{sec:implementation}, one challenge  to
specifying comprehensive rules for a MAC system that enforces
default-deny semantics is reducing the number of false negatives.  
In running our case studies, we found that 
the vast majority of false negatives would arise primarily due to Python's
internal library loading and DNS resolution processes.

To examine this challenge, we ran each case study application, 
in \system under AppArmor in complain mode: we identified a total of 42
common files that need to be read-accessible, plus 4 network protocols,
for the Python interpreter to run unimpeded, and for the applications
to connect to remote network destinations. 
%Our prototype's reliance on AppArmor
%requires us to specify a separate rule for each file and network protocol.
%The resulting AppArmor policy rules are available in our codebase.

In addition, each application required
explicit access to its parent directory as well as all imported Python modules 
contained within, which corresponds to up to 6 additional access rules for
the \texttt{alexa} application.
Since we cannot expect developers to manually specify around 50
access rules needed by their application by default, we developed a
policy generation tool that creates an access policy template
pre-populated with rules for the 46 common files and network protocols.
To further ease policy specification, our policy generation tool lists all files
in the application's parent directory and adds rules for all identified
files. Developers may then manually inspect the template and
modify any rules that are function-specific.

%Table~\ref{tab:app-policies} summarizes the number of
%application-specific access rules for each case study.
The majority of the manually added access rules for all three applications
are multiple network destination rules for the same library function.
%Two limitations of our \system prototype
%account for this. First, 
\system does not currently support
domain name-based access rules as this feature requires 
that all domain names be resolved a priori for \system's
IP-address based network destination verification. 
%As a result, developers must currently specify a separate network destination
%rule for each IP address prefix their application connects with.
Nonetheless, our case studies require no more than 11 function-level
network destination rules in the \texttt{plant\_watering} case.

%These numbers could be further reduced by adding support for rule grouping.
%That is, for resources that are accessible by multiple functions with the same
%access privileges, \system could support allowing developers to express
%these rules within a single rule. In the case of the \texttt{alexa} app, for instance, which
%currently requires 10 network access rules because two functions in the requests library
%connect to the same 5 network destinations, the number of required rules could be
%reduced by half with rule grouping, and even more with support for 
%domain name-based rules.
%Appendix~\ref{sec:case-study-pols} shows the function-level policy rules
%for each of our case studies.

%In summary, false negatives are the primary concern for \system's 
%policy specification. We believe that default rules, coupled with
%our policy generation tool, significantly reduce the number of false negatives
%application developers encounter.

\subsection{Vulnerability Analysis}
\label{secsec:sec-analysis}

To understand how effective \system's protections are 
against security vulnerabilities and exploits, we study 
\system's ability to mitigate specific instances of reported
Python library vulnerabilities (recall~\S\ref{secsec:lib-vulns}).
%, and analyze
%\system's ability to mitigate these cases. 
We emphasize that all of our analyzed vulnerabilities
could affect \emph{any} IoT use case, so our choice for testing a particular
vulnerability in a specific application does not reflect prevalence of
a vulnerability class in that IoT use case.

Since most of the reported vulnerabilities
do not affect the libraries in our case studies,
we replicate all analyzed vulnerabilities in a specially crafted
adversarial library targeting our case studies. 
We then call individual functions in our library from
our case study applications.
%, and modify the corresponding access policies
%to allow our adversarial functions to access the necessary OS resources
%and network destinations. 

We place the 9 reported attack classes into three broad categories: (1) successfully 
mitigated vulnerabilities, (2) case-dependent
for vulnerabilities that \system may mitigate in some instances, and
(3) beyond scope for attacks that fall outside \system's threat model.

%\begin{table*}[t]
%\centering
%\caption[\system's applicability to reported and synthesized library vulnerabilities.]{\label{tab:cve-mitigation} \system's ability to mitigate reported and synthesized Python library vulnerabilities.
%}
%\begin{tabular}{ccc}
%\toprule
%\textbf{Attack Class} & \textbf{Applicability} & \textbf{Mitigated Instance} \\
%\midrule
%Symlink attack & mitigated &  \\
%Direct data leak & case-dependent & file- and socket-based \\
%Arbitrary code execution & case-dependent & shell injection \\
%MITM & beyond scope & \\
%Web attack & beyond scope & \\
%Denial of service & beyond scope & \\
%Authentication bypass & beyond scope &  \\
%Weak crypto & beyond scope & \\
%Replay/data spoofing & beyond scope &  \\
%\bottomrule
%\end{tabular}
%\end{table*}

\Paragraph{Direct data leaks.}
We analyze three distinct instances of OS resource-based data leaks,
in which code with insufficient privileges attempts to gain access to
a sensitive OS resource. To this end, we craft two adversarial functions,
which upload an SSH private key
instead of the authorized photo file using the legitimate
tweepy library call, and which upload the authorized
photo to an unauthorized remote server, respectively. 
In addition, we also replicate the data leak bug reported in 
CVE-2019-9948~\cite{urllib-cve}, 
in which the Python urllib HTTP library exposes an unsafe API that allows
arbitrary local file opens.

We test these vulnerabilities using the \texttt{twitterPhoto} app,
and find that \system can successfully mitigate these problems.
However, we classify direct data leaks mitigation as case-dependent 
because several reported data leak vulnerabilities arise
due to in-memory data exposures. In contrast, \system currently 
only ensures the protection
of the runtime call stack memory, but does not isolate sensitive application-level
in-memory data objects, or protect against control-flow attacks.

\Paragraph{Symlink attacks.} A small number of reported vulnerabilities in Python
libraries comprises symlink attacks, in which an adversary attempts to
access an unauthorized file via a specially crafted symbolic link. We analyze
this attack by crafting an adversarial function that attempts to open 
\texttt{plant\_watering}'s private key through a symlink in the 
\texttt{\textbackslash{}tmp} directory.
Since our prototype follows all accessed symbolic links
to their source,
\system detects the attempt to access an unauthorized file successfully
mitigating this type of attack.

\Paragraph{Arbitrary code execution.}
A number of reported Python library vulnerabilities pertain to shell injection
made possible due to unsanitized input. To examine this attack class,
we crafted an adversarial
library function that attempts to directly exec a shell command as part of the
\texttt{plant\_watering} moisture sensor reading (\ie random number generator) call.

\system successfully mitigates the exec as the implicit file open is blocked.
Thus, we believe that
our analysis demonstrates that \system could be effective
in mitigating all instances of such shell injection attacks as they 
all ultimately require access to an unauthorized executable binary file. 
On the other hand, \system does not mitigate buffer overflow-based arbitrary code
execution attacks.
% Such attacks typically leverage a buffer overflow vulnerability
%that allows an attacker to hijack the control flow of the victim application. 
Further research is necessary to determine if \system's memory domains could
mitigate these attacks.

\Paragraph{Beyond scope vulnerability classes.}
A large portion of reported Python library vulnerabilities are beyond
the scope of \system's protections.
The majority of MITM vulnerabilities in our CVE reports analysis stem from a failure to
verify TLS certificates. The weak crypto vulnerabilities in our survey
primarily arise because a known-weak algorithm or random number generator 
was used, or because input was not properly validated. Similarly, 
\system cannot prevent replay or data spoofing attacks as these stem from
improper input (\ie nonce or filename) validation as well.
Most of the authentication
bypass bugs occur in larger Python-based frameworks
that fail to properly implement authentication procedures. 

Interestingly, the majority of the reported DoS attacks in Python libraries
stem from improper input handling or memory management that can 
cause the application to crash. While \system does not verify the
correctness of application or library code, we see an avenue for
using \system to prevent network-based DoS attacks.

%\subsection{Limiting MAC Check Circumvention}
%\label{secsec:circumvention-analysis}
%
%Our analysis in ~\S\ref{secsec:sec-analysis} demonstrates that 
%\system can successfully mitigate 
%several classes of OS resource-based vulnerabilities.
%However, we also wanted to evaluate how robust \system is to attempts by
%more clever adversaries to circumvent our function-level MAC mechanism.
%As we describe in~\S\ref{secsec:stack-inspection}, a library function
%without privileges to access a given resource may attempt to 
%fool the MAC checks by calling a more privileged library function
%that does have sufficient privileges to access the desired resource.
%
%Thus, we analyzed two specific instances:
%(1) an entirely unprivileged
%library function (\ie it does not appear in any rules in the 
%application's access policy) calls an authorized function, and (2) a function
%authorized to access an OS resource calls a function with privileges to
%a different resource.
%Because \system's stack inspection mechanism in the kernel 
%always begins by inspecting the outermost caller (\ie the bottom of the
%stack), it will immediately return to the 
%a ``permission denied'' error to the application, as the stack inspector
%cannot find an ACL entry for the outermost function call in the stack.

\subsection{Dynamic Language Features}
\label{secsec:dyn-analysis}

Prior proposals have recognized the potential security threat posed by
dynamic language features such as reflection and native code 
execution~\cite{flexdroid,breakapp}. As we describe in~\S\ref{secsec:lib-vulns},
Python's dynamic features enabled us to replace function pointers (aka 
monkey patching), leak a sensitive file at import time, and 
modify contents of Python stack frames including the value of
function arguments. To understand how these dynamic language features
affect \system's protections, we analyzed these three cases.
%, as well as external binary execution.

Much as in the direct data leak attack analysis, we found that \system
readily prevented unauthorized file accesses at import time. 
However, we met several challenges to \system's ability to
prevent all forms of stack frame tampering and monkey patching.
We found that \system's stack memory domain prevents native code from
directly accessing arbitrary stack frame memory.
Yet, because Python stores the local variables for each stack frame
in a separate dictionary data structure, pointed to by
the stack frame, our implemented stack frame isolation
is insufficient to prevent tampering with function arguments by native code
or monkey patching. As part of ongoing research, we are exploring more
robust countermeasures to these dynamic language features. 
\section{Evaluation}
\label{sec:eval}

To evaluate the performance of \system, we ran our three 
case studies (\S\ref{secsec:apps}) in vanilla Python and 
\system Python, measuring the execution time and memory overheads. We also
took microbenchmarks of the main Pyronia operations, as well as 
common system calls used in IoT applications 
to analyze the impact of \system's call stack inspection-based access control. 

Our testing system is an Ubuntu 18.04 LTS virtual machine
running our instrumented \system Linux Kernel on a single Intel Core
i7-3770 CPU, with 1.95 GB of RAM. 
Though not a dedicated IoT platform, our test VM's configuration is
comparable to recent single-board computers targeting IoT, 
such as the Raspberry Pi 4~\cite{rpi4} or the NVIDIA Jetson Nano~\cite{nvidia-jetson}.

To ensure that the results of our evaluation are consistent, we make minor modifications 
to our case study applications replacing their real-time data collection (\eg reading an image from a
camera) with a static data source (\eg an image file), and run
the applications for a finite number of iterations.
We emphasize that none of these modifications affected policy specification, 
or were needed to add support for \system's security mechanisms.

\subsection{Execution Time Overhead}
\label{secsec:time-overhead}

%\begin{figure}[t]
%  \centering
%    \includegraphics[width=0.47\textwidth]{diagrams/connect-latency}
%    \caption{\label{fig:connect-micro-latency} Mean latency in microseconds for the Python socket connect call with varying call stack depths. }
%\end{figure}

To analyze the impact of \system on application execution time, we measured 25 runs of
the end-to-end execution time for a single iteration, 
as well as the per-iteration execution time over 100 iterations. 
The measurement for a single iteration of the application
represents the worst-case scenario as it includes any
overhead due to \system initialization (and teardown),
and the kernel's call stack log is empty.
Measuring the per-iteration execution time, on the other hand, gives an estimate of the
long-term operating time of the application, and the overhead due to \system's run-time security checks, \ie call stack inspection and stack domain-related operations.

%As a baseline for \system's execution time overhead, we ran a set of measurements
%with a default access policy that does not restrict access to sensitive system resources 
%to specific library function calls. In this setup, \system still interposes on all system calls,
%but skips stack inspection entirely. 
%In the full \system experiment with library
%function-specific access rules, we measure the worst case execution time, in 
%which the entire call stack must be inspected upon a system call.

%\begin{table}[t]
%\centering
%\caption[Mean \system execution time overheads.]{\label{tab:pyr-latency-overheads} Mean execution time overhead for the first app iteration (end-to-end), and long-term execution.}
%\begin{tabular}{cccc}
%\toprule
% & \textbf{end-to-end} & \textbf{long-term} \\
%\hline
%\texttt{twitterPhoto} & 5x & 3x \\
%\texttt{alexa} & 5x & 2x \\
%\texttt{plant\_watering} & 2x & 3x \\
%\bottomrule
%\end{tabular}
%\end{table}

%Table~\ref{tab:pyr-latency-overheads} shows the mean end-to-end execution
%time overhead over 25 runs of a single iteration of our tested applications.
While \system's mean end-to-end execution time overhead of 2-5x is significant,
the mean long-term overhead per iteration is reduced to 2-3x.
Nonetheless, in absolute terms, the worst-case execution time for a 
single app iteration under
\system is 1.5 seconds for the \texttt{twitterPhoto} app, which we expect 
would remain largely imperceptible to end users in real-world 
deployments. 
%Further, these overheads could be counteracted by adjusting the
%loop interval (\ie the sleep time) of these applications based on the overhead imposed by 
%\system. 

\begin{figure}[t]
  \centering
    \includegraphics[width=0.47\textwidth]{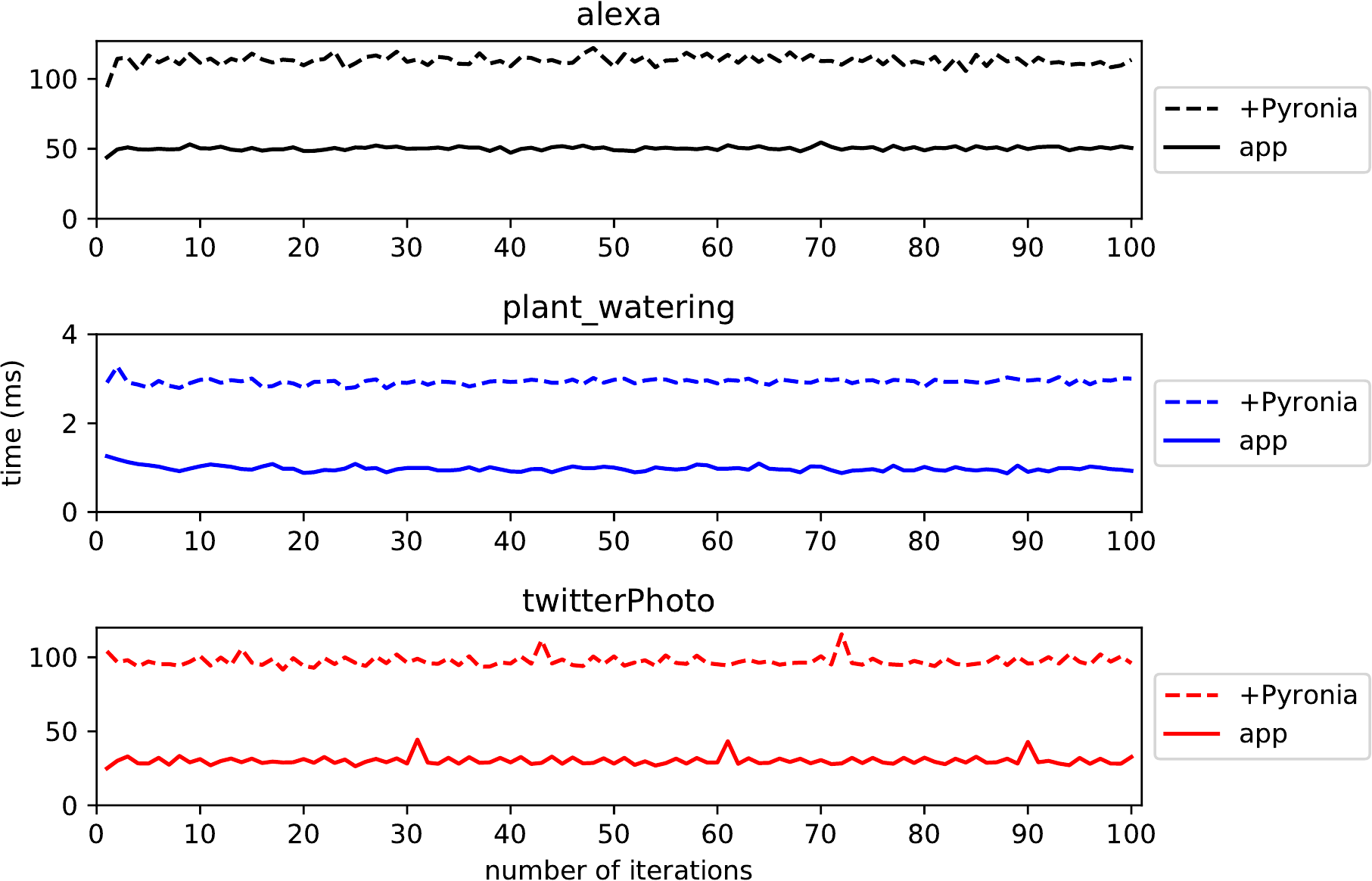}
    \caption[Mean \system per-iteration execution time.]{\label{fig:apps-iter-latency} Mean per-iteration execution time in seconds for each application with and without \system enabled.}
\end{figure}

\begin{figure}[t]
  \centering
    \includegraphics[width=0.47\textwidth]{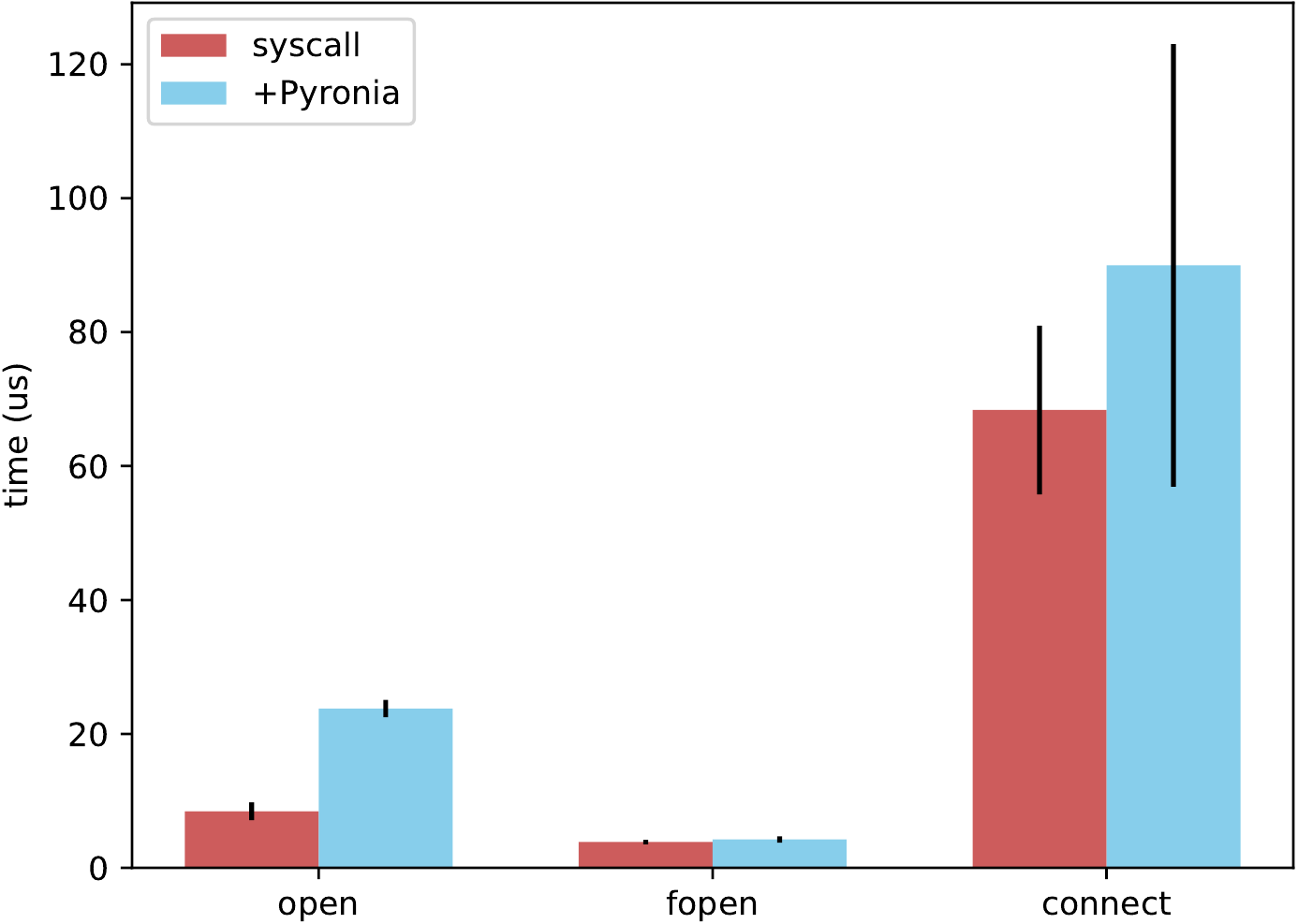}
    \caption[\system syscall microbenchmarks.]{\label{fig:pyr-syscall-latency} Mean execution time for 
    open, fopen and socket connect system calls across all tested applications.}
\end{figure}

Figure~\ref{fig:apps-iter-latency} plots the mean per-iteration execution time
over 5 runs of 100 iterations of our tested applications. Despite the overall
execution time overhead of \system, we observe that stack logging
lowers the long-term overhead of the \texttt{plant\_watering} and \texttt{twitterPhoto}
apps, as the execution time mostly levels after about 5 iterations.

Nonetheless, stack logging seems to play a little role in reducing the
execution time overhead for the \texttt{alexa} app.
%, as the end-to-end and long-term
%per-iteration overhead is 2x in both cases. 
For an application that requires active user involvement, 
\system's long-term overhead of 2x for
the \texttt{alexa} app
%compared to an original execution time of about 140 ms,
is likely unacceptable in a real-world deployment, even with an
absolute per-iteration execution time of under one-eighth of a second (113.3 ms).

\Paragraph{\system operation microbenchmarks.} 
Measurements of nine key Pyronia operations
%over 25 runs of a single iteration of our case studies
show that stack domain dynamic permissions adjustments greatly dominate the
overall \system overhead, with millions of domain access grant/revoke
calls in a single iteration in all applications. 
By comparison, the median number of all stack-related operations, 
\ie stack collection and hashing, is only in the teens. 
Thus, we attribute the main source of \system's run-time
overheads to stack domain access grant calls recorded in our experiments.
%Performance improvements for interactive IoT applications remain as future work.

%\Paragraph{\system operation microbenchmarks.} 
%To gain a better understanding of the source of \system's overheads
%we ran microbenchmarks of 9 main Pyronia API operations over 25 runs of a 
%single iteration of our case studies. Our results show that 
%stack domain dynamic access adjustments greatly dominate 
%the \system overhead, with the median number of access grant calls 
%of over 1 million for all tested applications. The significantly higher 
%execution time for the \texttt{alexa} app can perhaps be attributed to the over 9 million
%interpreter memory domain access grant calls recorded in our experiments.
%Performance improvements for interactive IoT devices remain as future work.

\Paragraph{Access control overhead.} 
%Our analysis of end-to-end
%and long-term per-iteration shows that the main module execution
%dominates the total execution time. 
%In other words, the 
%main contributor to \system's overhead are the run-time access checks
%and memory domain  the system call interposition and call stack 
%inspection needed to enforce \system's library function-level access policies. 
To further characterize the performance costs due to \system's access control
checks in the kernel, we ran microbenchmarks of the libc \emph{open()}, \emph{fopen()}
and \emph{connect()} (and their 64-bit variants), for which 
we have implemented our stack logging optimization. 
Figure~\ref{fig:pyr-syscall-latency} shows the mean 
execution time over 25 runs of a single iteration of our tested apps.
Our results %are consistent with our Pyronia operation microbenchmarks discussed 
%above, 
show that \system's system call interposition imposes at most
a 2x overhead for the \emph{open()} system call. 
%Nonetheless, in absolute terms, the overhead is 
%reasonable: the time to open a file or 
%connect to a socket only reaches 1ms (still a modest number) when the call stack is 
%50 levels and 100 levels deep, respectively. 
%Further, And (per~\S\ref{sec:app-analysis} and our benchmarks), the 
%median library call stack depth for the top-50 Python libraries is under 30 levels, and
%no greater than 14 levels for our tested applications.

\Paragraph{Summary.}
\system's execution time overhead is not trivial, despite
our performance optimizations. 
While some additional
time is spent during each system call, 
the main slowdown occurs due to dynamic memory domain 
page access adjustments. 
%module execution. We attribute this
%to two factors: (1) \system's access checks, and (2) memory domain access privilege
%adjustments to support Python's garbage collection.
Nonetheless, because the majority of IoT applications run on devices
only passively collecting and transmitting data, we expect these
overheads would go largely unnoticed by end users.
We plan to investigate further performance optimizations of memory
domain access adjustments, especially for interactive
IoT applications, as part of future work.

%, so the number of nested 
%functions calls needed to reach an overhead of 1ms far exceeds the call stack depths 
%we observe in real applications. 
%Thus, in real application's \system's overhead on these applications is acceptable.

%\begin{table*}[t]
%\centering
%\begin{tabular}{cccccccc}
%\toprule
% & \textbf{total \# dom pages} & \textbf{\# interpreter dom pages} & \textbf{min} & \textbf{median} & \textbf{mean} & \textbf{mode} & \textbf{max} \\
%\hline
%\texttt{hello} & 16 & 16 & 264 & 336.0 & 339.0 & 336.0 & 408 \\
%\texttt{twitterPhoto} & 125 & 120 & 240 & 336.0 & 360.4 & 336.0 & 1080 \\
%\texttt{alexa} & 92 & 85 & 168 & 336.0 & 345.1 & 336.0 & 816 \\
%\texttt{plant\_watering} & 42 & 42 & 240 & 336.0 & 336.0 & 336.0 & 408 \\
%\bottomrule
%\end{tabular}
%\begin{small}
%\caption{\label{tab:memdom-analysis} Memory domain metadata allocations in bytes under \system for each application. The number of stack domain pages includes the 
%domain pages reserved for the stack inspector thread.}
%\end{small}
%\end{table*}

\subsection{Memory Overhead}
\label{secsec:mem-overhead}

%\begin{figure}[t]
%  \centering
%    \includegraphics[width=0.47\textwidth]{diagrams/memusage}
%    \caption{\label{fig:memusage} Median memory usage in MB over time (in seconds) for each application with and without \system enabled.}
%\end{figure}

\system imposes memory overhead due to the creation
and management of the stack memory domain.
To evaluate the impact of this domain on userspace
memory consumption,
%we first measure the memory footprint of the metadata
%for the stack domain.
%as the language runtime maintains memory domains for the security-critical interpreter
%state (\ie call stack and stack inspector), and for data object isolation.
%To evaluate the impact memory domains alone have on memory consumption, we 
%first measure how much additional memory
%\system uses to maintain each domain. 
%Recall from~\S\ref{secsec:kernel} that
%each memory domain consists of a set of domain pages. 
%Thus, depending on the complexity of the
%application, and the developer's use of data object domains for additional protection,
%the number of domain pages required for a given memory domain will vary.
%Specifically, 
%For a more accurate estimate of \system's memory overhead due to memory domains, 
we first measure the userspace \emph{per-domain page}
metadata allocations, \ie the memory required for the \system runtime to maintain 
each domain page and the associated memory management data structures for the stack domain.
%for individual allocated blocks within a page
\footnote{We do not evaluate the actual \emph{data} allocation
overhead per domain page as \system does not change the amount of data 
the runtime allocates, only where in the runtime's address space this data is placed.}

Our analysis shows that the mean per-domain page metadata memory usage
for all tested applications is between 0.31 and 0.38 KB; the fact that
the page metadata allocations varies this little across 
all tested applications demonstrates that
the majority of domain pages contain a similar 
number of allocated blocks (\ie runtime stack frames), 
regardless of the total number of allocated domain pages.
Table~\ref{tab:memdom-analysis} further shows that the median memory usage of the
whole \system subs-system in the Python runtime remains under 200 KB,
even for applications with over 100 stack domain
pages.
These results are a strong indication that \system's domain
memory consumption scales linearly with the number of allocated domain pages.

\begin{table}[t]
\centering
\caption[\system per-domain page memory usage.]{\label{tab:memdom-analysis} Mean \system domain metadata memory usage.}
\begin{tabular}{ccc}
\toprule
 & \textbf{\# dom pages} & \textbf{\system total} \\
\hline
%\texttt{hello} & 17 & 22.3 KB \\
\texttt{twitterPhoto} & 176 & 151.2 KB \\
\texttt{alexa} & 141 & 147.9 KB \\
\texttt{plant\_watering} & 54 & 68.4 KB  \\
\bottomrule
\end{tabular}
\end{table}

\begin{table}[t]
\centering
\caption[Peak \system memory overhead.]{\label{tab:mem-usage} Peak memory overhead under \system.}
\begin{tabular}{ccc}
\toprule
 & \textbf{peak usage (in MB)} & \textbf{overhead} \\
\hline
\texttt{twitterPhoto} & 40.9 & 12.9\% \\
\texttt{alexa} & 33.9 & 33.0\%  \\
\texttt{plant\_watering} & 26.4 & 70.0\% \\
\bottomrule
\end{tabular}
\end{table}

%Our analysis also shows that the \system sub-system in 
%the Python runtime requires no more than a few
%hundred kilobytes of additional RAM for all tested applications (see
%Table~\ref{tab:memdom-analysis}).
Furthermore, \system's memory domains have a small impact on the overall memory 
consumption of our tested applications.
Table~\ref{tab:mem-usage} shows the median peak memory usage and overhead
over 5 100-iteration runs.
For the \texttt{twitterPhoto} application with a peak memory usage
of about 40 MB, \system's memory overhead is only 12.9\%, 
even with the largest number stack domain pages. 

\Paragraph{Summary.} 
\system incurs low memory overhead, even for
IoT applications that allocate over 100 domain pages.
For instance, domain metadata only consumes a total of 
151 KB for the \texttt{twitterPhoto} application with
176 allocated domain pages. For applications with
a greater number of domain pages, our results indicate that
the metadata memory overhead would likely grow linearly.
While the increase to
application-wide memory usage is the highest for the 
\texttt{plant\_watering} application at 70\%,
a peak memory consumption of under 30 MB is still rather modest.
Therefore, \system's memory overhead would not place an excessive 
burden on IoT devices with more constrained resources
than our testing system.

%Despite efforts, compatibility issues required us to run
%the \texttt{plant\_watering} application under the \texttt{memory-profiler} with 
%a default policy (\ie without \system's stack inspection), so the memory overhead of
%28.1\% (or 4.1 MB) we observe is likely an under-estimate. Yet with a 1.3x increase in 
%memory consumption with only two more data object domain pages than \texttt{twitterPhoto} 
%(with a 1.0x memory overhead), \system's memory overhead would still not place 
%an excessive burden.

\section{Discussion}
\label{sec:discussion}

\Paragraph{Multi-threading.}
While \system currently targets single-threaded
IoT applications, we discovered in~\S\ref{sec:app-analysis} as well
as during our experiments that 
a small number of IoT applications and libraries (about 10\%) spawn
pthreads.
This programming pattern introduces one key security challenge:
since threads execute 
independently of the main thread. In other words, a
vulnerability could still cause a confused deputy attack (violating~\textbf{P2}).

To address this issue, upon \texttt{pthread\_create()} or \texttt{thread\_start()} calls,
\system could automatically %inspect the state of the runtime call stack at that time,
save the state of the ``parent'' call stack, so as to provide the Access Control module with the
full provenance when the child thread makes a system call. Nonetheless, 
accurately mapping ``parent'' stacks to child stacks,
especially in the scenario of nested multi-threading, would be an additional
design challenge.

%\Paragraph{Preventing \system API misuse.} 
%\system currently gives the language runtime
%full authority over security-sensitive operations (\eg memory
%domain allocations and privilege adjustments, or call stack generation),
%and trusts that only the runtime is making such \system API calls.
%However, a resourceful adversary could use the API 
%to bypass \system's protections, or otherwise do harm, for instance
%by freeing domain-protected memory.
%
%To prevent such misuses of the \system API, we envision two potential
%approaches. In the style of~\cite{quire}, \system could use cryptographic
%mechanisms to authenticate the Netlink messages the runtime (including
%the stack inspector thread) sends to
%the kernel. Another approach would be use static analysis to identify
%rogue \system API calls in third-party code prior to launching the runtime.
%Further investigation is required to determine which method is more suitable.

\Paragraph{Improving policy specification.}
As we discuss in~\S\ref{secsec:policy-usability}, 
\system aims to reduce the burden of defining fine-grained access policies,
and lower false negatives.
However, due to our reliance on AppArmor, \system currently expects 
path-based access rules, which are often difficult to determine for
resources such as sensors. %, likely still places an undue burden on developers.

Designing a more developer-friendly and rigorous policy specification model is 
beyond the scope of \system.
One interesting approach may be to support simple mobile-style resource access capabilities
(\eg \texttt{READ\_CAM}), which \system could then automatically map
to the corresponding low-level system resources.

More short-term improvements include adding support for domain-based
network whitelisting, and maintaining a list of the most
critical default rules within the kernel,
allowing developers to remain agnostic to the runtime defaults.
The number of required rules could be further reduced by 
adding support for rule grouping.
That is, for resources accessible by multiple functions with the same
privileges, \system could support allowing developers to express
these policies as a single rule. %In the case of the \texttt{alexa} app, for instance, which
%currently requires 10 network access rules because two functions in the requests library
%connect to the same 5 network destinations, the number of required rules could be
%reduced by half with rule grouping, and even more with support for 
%domain name-based rules.

While we believe that the risks of completely automated
policy generation (\eg as in~\cite{systrace, passe})
outweigh the benefits, we see an opportunity for the
library developer community to ease the policy specification
process further. For instance, library developers could
contribute resource ``manifests'', \ie a list of required files and network 
destinations, and package these manifests along with their source 
code or binaries.
With support from \system, application developers could then automatically
load these manifests as part of their application-specific access policy,
allowing application developers to focus on their high-level policy.

\section{Conclusion}
\label{sec:conclusion}

We have presented \system, an intra-process
access control system for IoT device applications
written in high-level languages.
\system enforces function-granular MAC
of third-party code via a three-pronged approach: 
system call interposition, stack inspection,
and memory domains.
Unlike prior approaches, \system runs unmodified applications,
and does not require unintuitive policy specification. 
We implement a \system kernel and \system Python runtime.
Our evaluation of three open-source Python IoT applications demonstrates
that \system mitigates OS resource-based data leak vulnerabilities,
and shows that \system's performance overheads are acceptable for
the most common types of IoT applications.

% use section* for acknowledgement
%\section*{Acknowledgments}

\bibliographystyle{abbrv}
\bibliography{main}

%\clearpage
%\onecolumn
\appendix
\section{CVE Reports for Python Libraries}
\label{sec:cve-list}

Our analysis of reported Python library vulnerabilities in~\S\ref{secsec:lib-vulns} covers Common Vulnerabilities
and Exposures (CVE) reports made between January 2012 and March 2019.
We found 123 reports for a total of 78 different Python libraries and frameworks
in this seven-year time frame, and we identified nine main attack classes. 
Table~\ref{tab:full-cve-list} shows the attack class and affected Python library 
or framework for each analyzed CVE report.

\begin{onecolumn}

\begin{longtable}{ccc}
\caption[Python library CVE Reports.]{Reported Python library vulnerabilities between Feb 2012 and June 2019.}
\label{tab:full-cve-list}\\
\toprule
\textbf{CVE Report} & \textbf{Vulnerability Class} & \textbf{Affected Library/Framework} \\
\midrule
\endfirsthead
\multicolumn{3}{c}{\tablename\ \thetable\ Continued.}\\
\toprule
\textbf{CVE Report} & \textbf{Vulnerability Class} & \textbf{Affected Library/Framework} \\
\midrule
\endhead
CVE-2019-9948 & Direct data leak & urllib* \\
CVE-2019-9947 & Web attack & urllib* \\
CVE-2019-9740 & Web attack & urllib* \\
CVE-2019-7537 & Arbitrary code execution & donfig \\
CVE-2019-6690 & Direct data leak & python-gnupg \\
CVE-2019-5729 & MITM & splunk-sdk \\
CVE-2019-3575 & Arbitrary code execution & sqla-yaml-fixtures \\
CVE-2019-3558 & DoS & Facebook Thrift \\
CVE-2019-2435 & Direct data leak & Oracle MySQL Connectors \\
CVE-2019-13611 & Web attack & python-engineio \\
CVE-2019-12761 & Arbitrary code execution & pyXDG \\
CVE-2019-11324 & MITM & urllib* \\
CVE-2019-11236 & Web attack & urllib* \\
CVE-2018-5773 & Web attack & python-markdown2 \\
CVE-2018-20325 & Arbitrary code execution & definitions \\
CVE-2018-18074 & Direct data leak & requests \\
CVE-2018-17175 & Direct data leak & marshmallow \\
CVE-2018-10903 & Direct data leak & python-cryptography \\
CVE-2018-1000808 & DoS & pyopenssl \\
CVE-2018-1000807 & DoS & pyopenssl \\
CVE-2017-9807 & Arbitrary code execution & OpenWebif \\
CVE-2017-7235 & Arbitrary code execution & cloudflare-scrape \\
CVE-2017-3590 & Authentication bypass & MySQL \\
CVE-2017-2809 & Arbitrary code execution & ansible-vault \\
CVE-2017-2809 & Arbitrary code execution & tablib \\
CVE-2017-2592 & Direct data leak & python-oslo-middleware \\
CVE-2017-16764 & Arbitrary code execution & Django \\
CVE-2017-16763 & Arbitrary code execution & Confire \\
CVE-2017-16618 & Arbitrary code execution & OwlMixin \\
CVE-2017-16616 & Arbitrary code execution & PyAnyAPI \\
CVE-2017-16615 & Arbitrary code execution & MLAlchemy \\
CVE-2017-1002150 & Web attack & python-fedora \\
CVE-2017-1000433 & Authentication bypass & pysaml \\
CVE-2017-1000246 & Weak crypto & pysaml \\
CVE-2017-0906 & Web attack & recurly \\
CVE-2016-9910 & Web attack & html5lib \\
CVE-2016-9909 & Web attack & html5lib \\
CVE-2016-9015 & MITM & urllib* \\
CVE-2016-7036 & Weak crypto & python-jose \\
CVE-2016-5851 & Web attack & python-docx \\
CVE-2016-5699 & Web attack & urllib* \\
CVE-2016-5598 & Direct data leak & MySQL \\
CVE-2016-4972 & Arbitrary code execution & python-muranoclient \\
CVE-2016-2533 & DoS & PIL \\
CVE-2016-2166 & Weak crypto & Apache QPid Proton \\
CVE-2016-1494 & Weak crypto & python-rsa \\
CVE-2016-0772 & Weak crypto & smtplib \\
CVE-2015-7546 & Authentication bypass & python-keystoneclient \\
CVE-2015-5306 & Arbitrary code execution & ironic-inspector \\
CVE-2015-5242 & Arbitrary code execution & swiftonfile \\
CVE-2015-5159 & DoS & python-kdcproxy \\
CVE-2015-3220 & DoS & tlslite \\
CVE-2015-3206 & DoS & python-kerberos \\
CVE-2015-2674 & MITM & restkit \\
CVE-2015-2316 & DoS & Django \\
CVE-2015-1852 & MITM & python-keystoneclient \\
CVE-2015-1326 & Arbitrary code execution & python-dbusmock \\
CVE-2014-9365 & MITM & urllib* \\
CVE-2014-8165 & Arbitrary code execution & powerpc-utils-python \\
CVE-2014-7144 & MITM & python-keystoneclient \\
CVE-2014-4616 & Direct data leak & simplejson \\
CVE-2014-3995 & Web attack & Django \\
CVE-2014-3994 & Web attack & Django \\
CVE-2014-3598 & DoS & PIL \\
CVE-2014-3589 & DoS & PIL \\
CVE-2014-3539 & Arbitrary code execution & rope \\
CVE-2014-3146 & Web attack & lxml \\
CVE-2014-3137 & Authentication bypass & Bottle \\
CVE-2014-3007 & Arbitrary code execution & PIL \\
CVE-2014-1934 & Symlink attack & eyeD3 \\
CVE-2014-1933 & Symlink attack & PIL \\
CVE-2014-1932 & Symlink attack & PIL \\
CVE-2014-1929 & Arbitrary code execution & python-gnupg \\
CVE-2014-1928 & Arbitrary code execution & python-gnupg \\
CVE-2014-1927 & Arbitrary code execution & python-gnupg \\
CVE-2014-1839 & Symlink attack & logilab-common \\
CVE-2014-1838 & Symlink attack & logilab-common \\
CVE-2014-1830 & Direct data leak & requests \\
CVE-2014-1829 & Direct data leak & requests \\
CVE-2014-1624 & Symlink attack & python-xdg \\
CVE-2014-1604 & Data spoofing & python-rply \\
CVE-2014-0472 & Arbitrary code execution & Django \\
CVE-2014-0105 & Authentication bypass & python-keystoneclient \\
CVE-2013-7459 & Arbitrary code execution & PyCrypto \\
CVE-2013-7440 & MITM & ssl \\
CVE-2013-7323 & Arbitrary code execution & python-gnupg \\
CVE-2013-6491 & MITM & python-qpid \\
CVE-2013-6444 & MITM & pyWBEM \\
CVE-2013-6418 & MITM & pyWBEM \\
CVE-2013-6396 & MITM & python-swiftclient \\
CVE-2013-4482 & Authentication bypass & python-paste-script \\
CVE-2013-4347 & Weak crypto & python-oauth2 \\
CVE-2013-4346 & Auth token replay attack & python-oauth2 \\
CVE-2013-4238 & MITM & ssl \\
CVE-2013-4111 & MITM & python-glanceclient \\
CVE-2013-2191 & MITM & python-bugzilla \\
CVE-2013-2132 & DoS & pymongo \\
CVE-2013-2131 & DoS & python-rrdtool \\
CVE-2013-2104 & Auth token replay attack & python-keystoneclient \\
CVE-2013-2013 & Direct data leak & python-keystoneclient \\
CVE-2013-1909 & MITM & Apache QPid Proton \\
CVE-2013-1665 & Web attack & xml \\
CVE-2013-1664 & DoS & xml \\
CVE-2013-1445 & Weak crypto & PyCrypto \\
CVE-2013-1068 & Authentication bypass & python-nova, python-cinder \\
CVE-2012-5825 & MITM & Tweepy \\
CVE-2012-5822 & MITM & zamboni \\
CVE-2012-5563 & Authentication bypass & python-keystoneclient \\
CVE-2012-4571 & Weak crypto & python-keyring \\
CVE-2012-4520 & Web attack & Django \\
CVE-2012-4406 & Arbitrary code execution & python-swiftclient \\
CVE-2012-3533 & MITM & ovirt-engine-python-sdk \\
CVE-2012-3458 & Weak crypto & Beaker \\
CVE-2012-3444 & DoS & Django \\
CVE-2012-3443 & DoS & Django \\
CVE-2012-2921 & DoS & python-feedparser \\
CVE-2012-2417 & Weak crypto & PyCrypto \\
CVE-2012-2374 & Web attack & tornado \\
CVE-2012-2146 & Weak crypto & elixir \\
CVE-2012-1575 & Web attack & Cumin \\
CVE-2012-1502 & Arbitrary code execution & PyPAM \\
CVE-2012-1176 & DoS & PyFriBidi \\
CVE-2012-0878 & Authentication bypass & python-paste-script \\

\bottomrule
\end{longtable}

\end{onecolumn}

\end{document}